\documentclass[twoside,australian,sort&compress]{iopart}
\usepackage[T1]{fontenc}
\usepackage[utf8]{inputenc}
\usepackage{geometry}
\usepackage{graphicx}
\usepackage[numbers]{natbib}

\makeatletter
\usepackage{iopams}
\usepackage{setstack}

\newcommand{\eqref}[1]{(\ref{#1})}

\usepackage{xurl}

\makeatother

\usepackage{babel}
\begin{document}
\title[]{Toward a complete and comprehensive cross section database for electron scattering from NO using machine learning}
\author{{\Large{}P W Stokes$^{1,2}$, R D White$^{1}$, L Campbell$^{3}$
and M J Brunger$^{3,4}$}}
\address{{\Large{}$^{1}$}{\large{}College of Science and Engineering, James
Cook University, Townsville, QLD 4811, Australia}}
\address{{\Large{}$^{2}$}{\large{}Department of Medical Physics, Townsville
University Hospital, Townsville, QLD 4814, Australia}}
\address{{\Large{}$^{3}$}{\large{}College of Science and Engineering, Flinders
University, Bedford Park, SA 5042, Australia}}
\address{{\Large{}$^{4}$}{\large{}Department of Actuarial Science and Applied
Statistics, Faculty of Business and Management, UCSI University, Kuala
Lumpur 56000, Malaysia}}
\ead{{\large{}michael.brunger@}flinders.edu.au}
\begin{abstract}
We review experimental and theoretical cross sections for electron
scattering in nitric oxide (NO) and form a comprehensive set of plausible
cross sections. To assess the accuracy and self-consistency of our
set, we also review electron swarm transport coefficients in pure
NO and admixtures of NO in Ar, for which we perform a multi-term Boltzmann
equation analysis. We address observed discrepancies with these experimental
measurements by training an artificial neural network to solve the
inverse problem of unfolding the underlying electron-NO cross sections,
while using our initial cross section set as a base for this refinement.
In this way, we refine a suitable quasielastic momentum transfer cross
section, a dissociative electron attachment cross section and a neutral
dissociation cross section. We confirm that the resulting refined
cross section set has an improved agreement with the experimental
swarm data over that achieved with our initial set. We also use our
refined data base to calculate electron transport coefficients in
NO, across a large range of density-reduced electric fields from 0.003
Td to 10,000 Td.
\end{abstract}
\noindent{\it Keywords\/}: {nitric oxide, electron scattering cross sections, electron transport,
machine learning\\
}
\submitto{\JCP}
\maketitle

\section{\label{sec:Introduction}Introduction}

Plasma medicine is a relatively new field that employs low-temperature
atmospheric pressure plasmas (LTAPP), in order to induce beneficial
effects in biological tissue \citep{Kong2009,Samukawa2012,Bruggeman2016,Adamovich2017}.
Key to these benefits is the formation and subsequent synergistic
interactions of reactive oxygen and nitrogen species (RONS) \citep{Kong2009}.
The accurate modelling, control, and optimisation of LTAPP for plasma
medicine applications is dependent on a complete and detailed understanding
of all plasma-tissue interactions, of which one important subset is
the interaction of electrons with RONS \citep{Samukawa2012,Bruggeman2016,Adamovich2017}.
One of the most important RONS is nitric oxide (NO) \citep{Kong2009},
which has been identified as being effective in both tissue disinfection
\citep{Shekhter2005} and apoptosis of cancer cells \citep{Kim2016,Li2017}.
A precise description of the interactions between electrons and NO,
in the form of electron-impact cross sections \citep{Tanaka2016},
is thus important for the predictive understanding of many plasma
treatments. With this motivation in mind, in this investigation, we
compile a comprehensive set of electron-NO cross sections that attempts
to address the shortcomings of those already available in the literature.

To ensure the accuracy and self-consistency of our electron-NO cross
section set, we calculate corresponding electron swarm transport coefficients
for comparison with measurements in the literature. Any discrepancies
observed in the swarm parameters can then be addressed through appropriate
refinements of the cross section set. This \textit{inverse swarm problem}
of unfolding cross sections from swarm measurements has a long and
successful history \citep{Mayer1921,Ramsauer1921,Townsend1922,Frost1962,Engelhardt1963,Engelhardt1964,Hake1967,Phelps1968}.
Nevertheless, when the amount of available swarm data becomes limited,
the inverse swarm problem becomes ill-posed and is no longer guaranteed
to have a unique solution. Traditionally, this difficulty was overcome
by relying on an expert in swarm analysis to rule out unphysical solutions
and to select the solution that is the most physically plausible.
Recently, however, Stokes \textit{et al}. were able to avoid this
time-consuming and possibly subjective process of manual iterative
refinement by employing an artificial neural network model \citep{Stokes2019},
to solve the inverse swarm problem for the biomolecule analogues tetrahydrofuran
(THF, $\mathrm{C}_{4}\mathrm{H}_{8}\mathrm{O}$) \citep{Stokes2020}
and $\alpha$-tetrahydrofurfuryl alcohol (THFA, $\mathrm{C_{5}H_{10}O_{2}}$)
\citep{Stokes2021}, with the former result being found to be of comparable
quality to a conventional refinement ``by hand'' \citep{DeUrquijo2019a}.
This application of machine learning was originally proposed by Morgan
\citep{Morgan1991} in the early 1990s. Given that machine learning
has in recent years been applied to problems from all fields of science,
including chemical physics \citep{Ceriotti2021}, revisiting the proposal
of Morgan with a modern machine learning methodology and hardware
was certainly overdue. However, most pivotal to the procedure of Stokes
\textit{et al}. was their use of the LXCat project \citep{Pancheshnyi2012,Pitchford2017,Carbone2021},
allowing for their neural network to be trained on a wealth of ``real''
cross sections, thus allowing their model to ``learn'' what constitutes
a physically-plausible cross section set. In this work, we employ
a similar machine learning approach to automatically and objectively
refine electron-NO cross sections given independent experimental swarm
data.

The remainder of this paper is structured as follows. In Sec. \ref{sec:Initial-cross-section},
we review relevant experimental and theoretical cross sections and
assemble our initial electron-NO cross section database. In Sec. \ref{sec:Multi-term-Boltzmann-equation},
we assess the accuracy and self-consistency of our initial set by
simulating corresponding transport coefficients, using a multi-term
Boltzmann equation solver, and comparing them to experimental measurements
present in the literature. Sec. \ref{sec:Cross-section-refinement}
details our data-driven approach to the swarm analysis of this experimental
data and presents the resulting refinements to our initial cross section
set. Sec. \ref{sec:Multi-term-Boltzmann-equation-refined} assesses
to what extent our refined set improves the agreement with experimental
swarm data over what we proposed initially. We also use our refined
set here to calculate transport coefficients for electrons in NO across
a large range of density-reduced electric fields. Finally, Sec. \ref{sec:Conclusion}
presents our conclusions and makes some suggestions for future work.

\section{\label{sec:Initial-cross-section}Initial cross section compilation}

\begin{figure}
\begin{centering}
\includegraphics[scale=0.7]{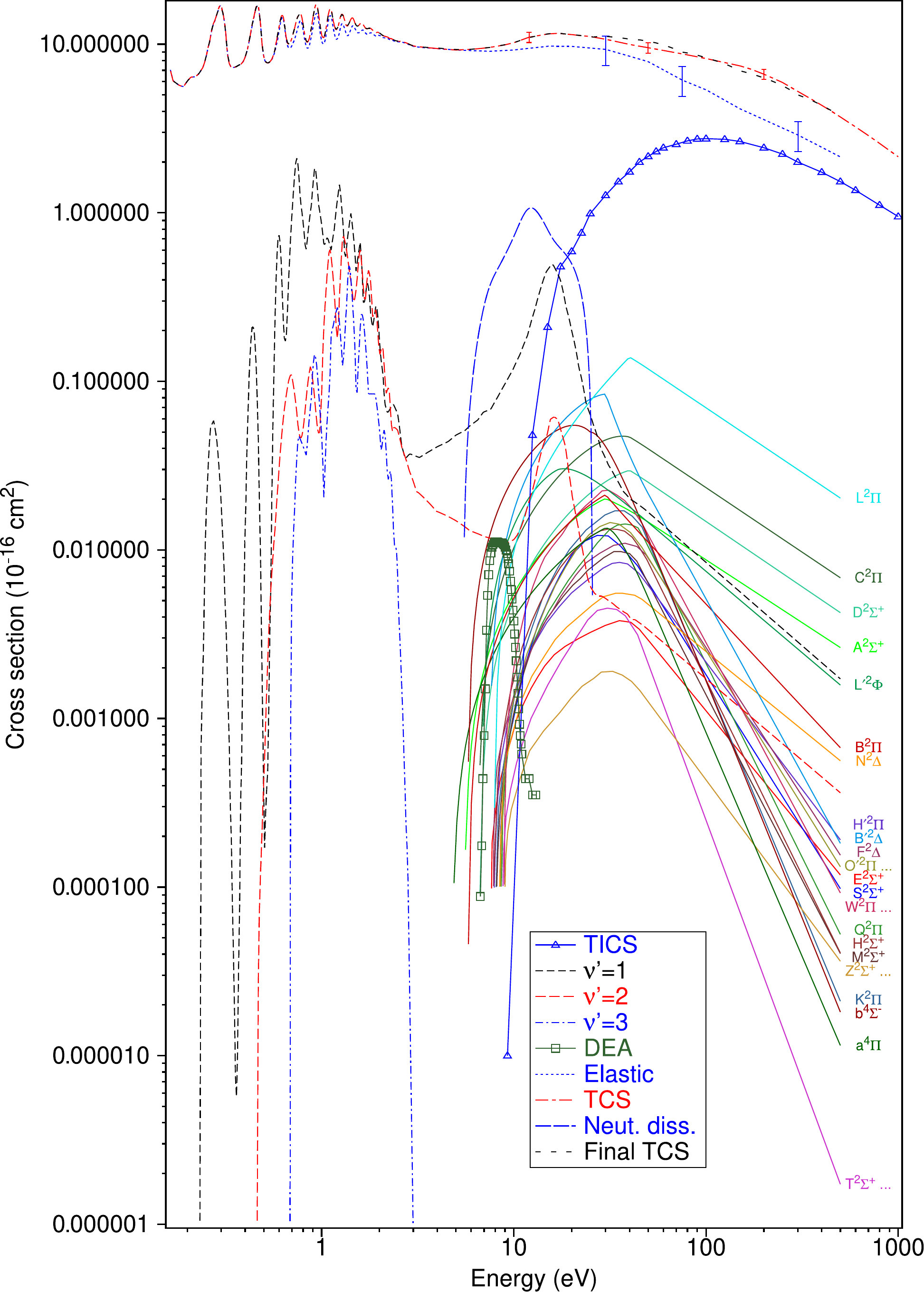}
\par\end{centering}
\caption{\label{fig:Graphical-representation-for}Graphical representation
for the initial cross section data base that we have assembled as
a part of the present study. See also our text and the legend in the
figure.}
\end{figure}
There have been three comprehensive attempts to compile cross section
data for electron scattering from NO. The first, by Brunger and Buckman
\citep{Brunger2002}, is now quite dated and so does not figure in
what follows. The second was by Professor Y. Itikawa \citep{Itikawa2016},
perhaps the doyen of scientists who collect and evaluate cross section
data, while the most recent was from Song \textit{et al}. \citep{Song2019}.
In his compilation, Itikawa \citep{Itikawa2016} reported data for
the grand total cross section (TCS), elastic integral cross section
(ICS), elastic momentum transfer cross section (MTCS), vibrational
excitation ICS for the $0\rightarrow1$, $0\rightarrow2$ and $0\rightarrow3$
quanta, a subset of the electronic-state ICS, a total ionisation cross
section (TICS) and a dissociative electron attachment (DEA) cross
section. Unfortunately, in many cases his recommendations did not
cover a broad enough energy range, nor were all the possible open
channels considered (neutral dissociation {[}ND{]} was not addressed),
to fulfil the criteria of Tanaka \textit{et al}. \citep{Tanaka2016}
and Brunger \citep{Brunger2017} that data bases for modelling and
simulation studies (e.g. \citep{Sanz2012,Brunger2016,White2018,DeUrquijo2019a})
must be comprehensive and complete. Song \textit{et al}. \citep{Song2019},
where possible, updated the Itikawa data base and then used that as
a starting point to solve the inverse electron-swarm problem \citep{Stokes2021}
in NO. In that approach, ICSs are varied, in conjunction with a two-term
approximation Boltzmann equation solver \citep{Pitchford2017}, in
order to force agreement between the simulated and measured transport
coefficient data \citep{Song2019}. The so-determined ICS set then
became their recommended cross section data base for the $e+\mathrm{NO}$
collision system. We have several reservations with the work of Song
\textit{et al}. \citep{Song2019}. Firstly, there is no \textit{a
priori} rationale provided by Song \textit{et al}. \citep{Song2019}
for why the two-term approximation to solving Boltzmann's equation
should be valid for all the transport coefficients they simulated
over the range of $E/n_{0}$ ($E$ = applied electric field strength;
$n_{0}$ = number density of the background gas {[}i.e. NO here{]}
through which the electron swarm drifts and diffuses) they considered.
Indeed, in Sec \ref{sec:Multi-term-Boltzmann-equation}, we apply
a multi-term Boltzmann solver and determine that the two-term approximation
has not converged sufficiently for some of the measurements considered
in their swarm analysis. Secondly, as part of their swarm analysis,
Song \textit{et al}. \citep{Song2019} use the swarm data of Takeuchi
\textit{et al}. \citep{Takeuchi2001} who determined their transport
coefficients from measured distributions of electron arrival times
\citep{Takeuchi2001}. It is known that these resulting ``arrival
time spectra'' transport coefficients are distinct from the bulk
(centre-of-mass) transport coefficients provided by many Boltzmann
solvers \citep{Robson2011,Kondo1990} and it is not clear whether
Song \textit{et al}. \citep{Song2019} have taken this into account
when performing their swarm analysis. Finally, the ill-posed nature
of the inverse swarm problem \citep{Stokes2020} can potentially lead
to uniqueness issues in the cross sections derived. Namely, while
the ICSs determined by Song \textit{et al}. \citep{Song2019} do lead
to transport coefficients that are largely consistent with the available
swarm data they might not be the only set that does so. Under those
circumstances the Song \textit{et al}. \citep{Song2019} data base
might not be physical. As a consequence of those concerns, we assemble
our initial $e+\mathrm{NO}$ cross section data base by using extensively
the work of Itikawa \citep{Itikawa2016}.

Brunger \textit{et al}. \citep{Brunger2000} reported ICSs for 28
excited electronic states of NO, but only over the limited incident
electron energy range 15–50 eV. Cartwright \textit{et al}. \citep{Cartwright2000}
extended a subset of 9 of those cross sections, to their various threshold
energies and out to 500 eV, in order to study the excited-state densities
and band emissions of NO under auroral conditions. Here we extend
the remaining 19 excited electronic-state ICSs to their respective
threshold energies \citep{Campbell1997} and out to 500 eV, with all
of these electronic-state ICSs being plotted in Fig. \ref{fig:Graphical-representation-for}.
Note that Xu \textit{et al}. \citep{Xu2018} reported some BE$f$-scaling
\citep{Tanaka2016} theoretical results for the $A^{2}\Sigma^{+}$,
$C^{2}\Pi$ and $D^{2}\Sigma^{+}$ electronic states, and found good
agreement with the ICSs of Brunger \textit{et al}. \citep{Brunger2000}
for the $A$- and $C$- states. For the $D$-state, however, a discrepancy
was noted although that may reflect on the applicability of using
the BE$f$-scaling approach for an electronic state with such a small
optical oscillator strength \citep{Kato2007}. For vibrational excitation
we essentially adopt Figure 1 from Campbell \textit{et al}. \citep{Campbell2004},
which was constructed from the low-energy results of Josić \textit{et
al}. \citep{Josic2001} and Jelisavcic \textit{et al}. \citep{Jelisavcic2003}
and the higher-energy results of Mojarrabi \textit{et al}. \citep{Mojarrabi1995}.
All these ICSs, for the $\nu$ = 0–1, 0–2 and 0–3 vibrational quanta,
are plotted in Fig. \ref{fig:Graphical-representation-for}. The present
DEA cross sections were taken directly from Table 6 of Itikawa \citep{Itikawa2016},
being originally sourced from the work of Rapp and Briglia \citep{Rapp1965}
(see Fig. \ref{fig:Graphical-representation-for}). Similarly, the
present TICS was taken directly from Table 5 of Itikawa \citep{Itikawa2016},
with the origin of the cross sections in this case being from Lindsay
\textit{et al}. \citep{Lindsay2000} (again see Fig. \ref{fig:Graphical-representation-for}).
In the case of the elastic scattering, our initial data set is constructed
below 3.38 eV by subtracting the present vibrational ICSs of Fig.
\ref{fig:Graphical-representation-for} from the TCS measurement of
Alle \textit{et al}. \citep{Alle1996}. Above 3.38 eV we use the recommended
data from Table 2 of Itikawa \citep{Itikawa2016}. The present elastic
ICSs are also plotted in our Fig. \ref{fig:Graphical-representation-for}.
The TCSs of our initial cross section compilation were formed as follows.
Below an incident electron energy of 2.6 eV we used the values of
Alle \textit{et al}. \citep{Alle1996}, as digitised from their Figure
1, while for $E_{0}$ above 5 eV we employed the values from Table
1 in Itikawa \citep{Itikawa2016}. Between 2.6 eV and 5 eV we undertook
an interpolation that ensured that the TCSs of Alle \textit{et al}.
\citep{Alle1996} merged smoothly with those recommended by Itikawa
\citep{Itikawa2016}. The resultant TCS from this process is shown
pictorially in Fig. \ref{fig:Graphical-representation-for}. Finally,
there remains the neutral dissociation cross section. There are no
known measurements or calculations for the ND ICS in $e+\mathrm{NO}$
scattering, so our approach for estimating it was as follows. The
sum of the individual ICSs, for all open scattering channels at some
$E_{0}$, should be consistent with the TCS, at least to within the
uncertainties on the TCS. In this case, we found that when we summed
all our ICSs there was a quite narrow energy region ($\sim$ 5 – 20
eV) where it underestimated the TCS (to well outside the error on
our TCS). We therefore assigned that residual cross section strength
to be the ICS for ND, with our derived result again being plotted
in Fig. \ref{fig:Graphical-representation-for}.

Fig. \ref{fig:Graphical-representation-for} therefore summarises
all the ICSs and the TCS that constitute our initial cross section
compilation for input into our multi-term Boltzmann equation solver
\citep{Stokes2020,Stokes2021}, in order to determine transport coefficients
that can be compared to corresponding independent results from electron-swarm
experiments. The present error estimates on those ICSs and the TCS
largely reflect those given in the original papers, from which our
initial data base was derived. Typically, this would be $\sim\pm25\%$
on the elastic ICS, $\sim\pm30\%$ on the vibrational excitation ICS,
$\sim\pm7\%$ on the TICS, in the range of $\sim\pm25\%$–$70\%$
for the various electronic-state ICS, $\sim\pm30\%$ on the DEA ICS
and $\sim\pm80\%$ on the ND ICS that we have derived. The independent
TCS data set would have an error of $\sim\pm7\%$, a little more than
cited in the relevant papers as we have included a small additional
uncertainty due to the so-called ``forward angle scattering effect''
\citep{Brunger2017a} that was previously not allowed for. It is important
to remember that, when using this initial cross section compilation
in conjunction with a multi-term Boltzmann equation solver, the errors
on the ICSs will in principle necessarily lead to a band of allowed
solutions for the transport coefficients that are to be simulated.

Another observation we should make in respect to Fig. \ref{fig:Graphical-representation-for}
is that it is clear that the TCS we determine from adding up all our
various ICSs (denoted by - - - in black and called ``Final TCS'')
is entirely consistent with the TCS we derived from the independent
measurements available to us (denoted by — - — in red). While this
is a necessary condition for a cross section compilation to be plausibly
physical, as we saw in our work on THF \citep{Stokes2020} and THFA
\citep{Stokes2021} it by no means \textit{a priori} guarantees that
the simulated transport coefficients will reproduce the available
swarm data. Finally, it is also relevant to note here that all the
ICSs in the present compilation are rotationally averaged, which is
a direct result of the energy resolution of current-day electron spectrometers
not being narrow enough to resolve the various rotational lines \citep{Brunger2002}.

\section{\label{sec:Multi-term-Boltzmann-equation}Multi-term Boltzmann equation
analysis of our initial set}

\begin{figure}
\begin{centering}
\includegraphics[scale=0.45]{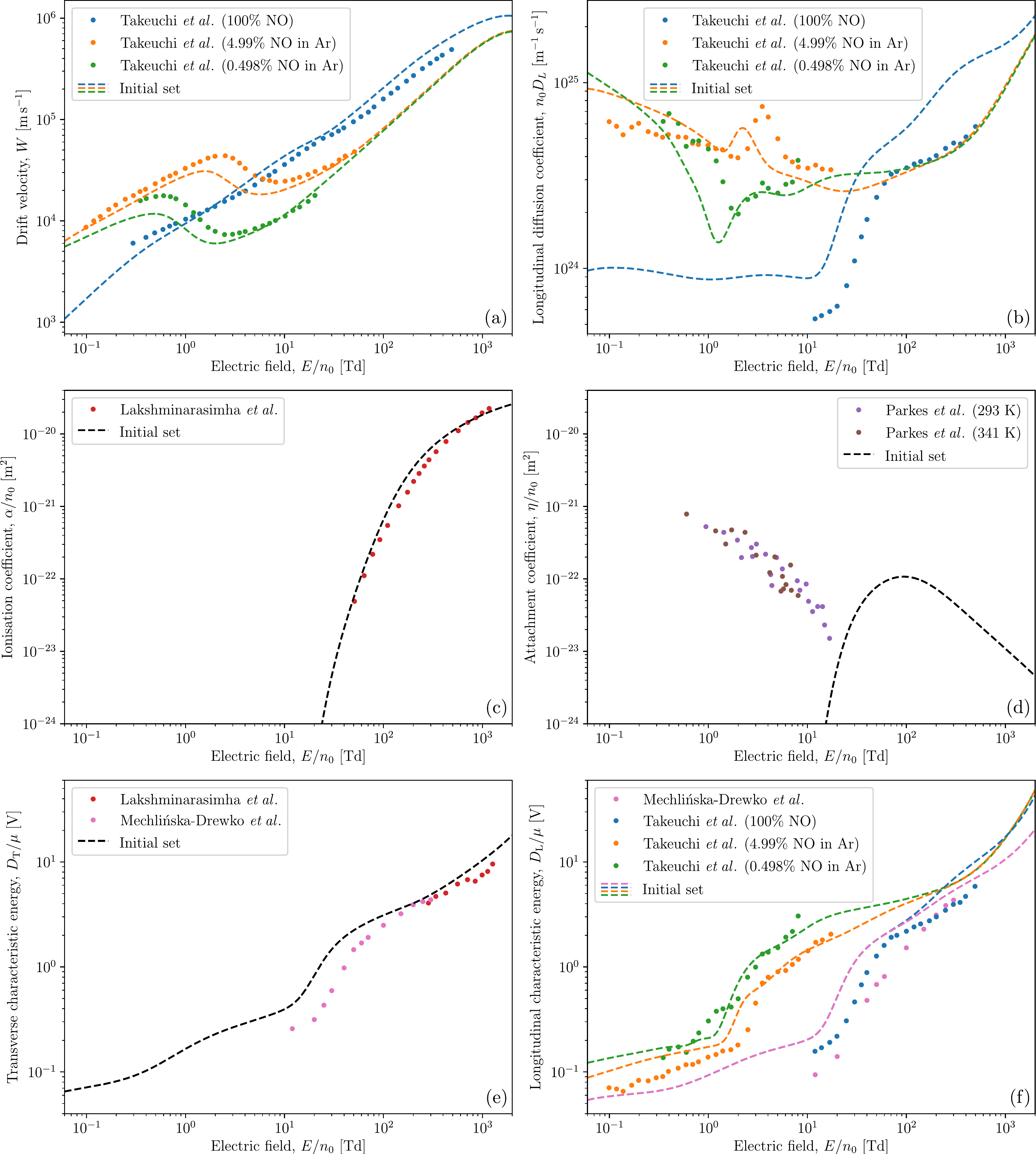}
\par\end{centering}
\caption{\label{fig:Comparison-of-transport}Transport coefficients calculated
for our initial electron-NO cross section set, alongside swarm measurements
from the literature for comparison. See also legends for further details.}
\end{figure}
To assess the quality of our initial set of electron-NO cross sections,
we apply a well-benchmarked multi-term solution of Boltzmann's equation
\citep{White2018,Boyle2017,White2003} in order to calculate swarm
transport coefficients for comparison to measured values in the literature.
Chronologically, electron-NO swarm measurements include drift velocities
and transverse characteristic energies by Skinker \textit{et al}.
\citep{Skinker1923} and Bailey \textit{et al}. \citep{Bailey1934},
transverse characteristic energies by Townsend \citep{Townsend1948},
drift velocities and attachment coefficients by Parkes \textit{et
al}. \citep{Parkes1972}, transverse characteristic energies and ionisation
coefficients by Lakshminarasimha \textit{et al}. \citep{Lakshminarasimha1977},
transverse and longitudinal characteristic energies by Mechlińska-Drewko
\textit{et al}. \citep{Mechlinska-Drewko1999}, and, most recently,
drift velocities and longitudinal diffusion coefficients by Takeuchi
\textit{et al}. \citep{Takeuchi2001} (for both pure NO and admixtures
of NO in Ar).

When performing a Boltzmann equation swarm analysis, it is essential
to precisely interpret what is being measured in each swarm experiment
under consideration \citep{Robson2011,Casey2020}. In this regard,
many of the earlier electron-NO measurements are inadequate as they
are obtained using a ``magnetic deflection'' experiment \citep{Townsend1913}
that measures a drift velocity that is distinct from the bulk (centre-of-mass)
drift velocity measured in pulsed-Townsend experiments \citep{Casey2020}
and provided by our Boltzmann solver. Ultimately, we choose to calculate
comparisons to the drift velocities and longitudinal diffusion coefficients
of Takeuchi \textit{et al}. \citep{Takeuchi2001}, the ionisation
coefficients of Lakshminarasimha \textit{et al}. \citep{Lakshminarasimha1977},
the attachment coefficients of Parkes \textit{et al}. \citep{Parkes1972},
the transverse characteristic energies of Lakshminarasimha \textit{et
al}. \citep{Lakshminarasimha1977} and Mechlińska-Drewko \textit{et
al}. \citep{Mechlinska-Drewko1999}, and the longitudinal characteristic
energies of Mechlińska-Drewko \textit{et al}. \citep{Mechlinska-Drewko1999}
and Takeuchi \textit{et al}. \citep{Takeuchi2001} (calculated using
their drift and diffusion measurements). We purposefully leave out
the 421 K measurements of Parkes \textit{et al}. \citep{Parkes1972},
due to an observed pressure dependence in their measured attachment
coefficients that they attribute to electron detachment from product
ions at this temperature. We also neglect the pure NO diffusion measurements
of Takeuchi \textit{et al}. \citep{Takeuchi2001} below 11 Td, due
to the pressure dependence arising from three-body attachment in this
low $E/n_{0}$ regime \citep{Bradbury1934,Parkes1972}. While their
associated drift velocity measurements also show a pressure dependence,
it is to a far lesser extent and Takeuchi \textit{et al}. \citep{Takeuchi2001}
are able to consistently estimate pressure-independent values by extrapolation
to zero pressure \citep{Takeuchi2001}. In almost all cases considered,
we interpret swarm measurements as pertaining to the centre of mass
of the swarm (i.e. bulk transport coefficients). The only exception
we make is for the double-shutter drift tube experiment of Takeuchi
\textit{et al}. \citep{Takeuchi2001}, which measures an ``arrival
time spectra'' drift velocity, $W_{m}$, that is distinct from the
bulk drift velocity when nonconservative effects are present \citep{Robson2011}.
Fortunately, we can approximate this quantity in terms of our simulated
bulk transport coefficients with \citep{Kondo1990}:
\begin{equation}
W_{m}\approx W-2\alpha_{\mathrm{eff}}D_{L},
\end{equation}
where $W$ is the bulk drift velocity, $\alpha_{\mathrm{eff}}$ is
the effective first Townsend ionisation coefficient, and $D_{L}$
is the bulk longitudinal diffusion coefficient.

To perform the transport calculations, we require an appropriate quasielastic
(elastic+rotation) momentum transfer cross section, which we obtain
by multiplying the present elastic ICS (from Fig. \ref{fig:Graphical-representation-for})
by the ratio of elastic MTCS to elastic ICS, each taken from the set
of Itikawa \citep{Itikawa2016}. When calculating the admixture transport
coefficients, we use the argon cross section set present in the Biagi
database \citep{Biagi,Biagib,LXCat}. The resulting simulated transport
coefficients are plotted alongside the aforementioned experimental
measurements in Figs. \ref{fig:Comparison-of-transport}(a)–(f) for
drift velocities, diffusion coefficients, ionisation coefficients,
attachment coefficients, transverse characteristic energies, and longitudinal
characteristic energies, respectively. Fig. \ref{fig:Comparison-of-transport}(a)
shows a qualitative agreement between our simulated drift velocities
and the measurements of Takeuchi \textit{et al}. \citep{Takeuchi2001}.
In the case of pure NO, our initial cross section set results in drift
velocities that are consistently too small below 2 Td and too large
above 2 Td. In the admixture cases, the onset of negative differential
conductivity (NDC) — the phenomenon of \textit{decreasing} drift velocity
with \textit{increasing} $E/n_{0}$ — occurs earlier in our simulations
than is shown by experiment. Fig. \ref{fig:Comparison-of-transport}(b)
indicates that our calculated longitudinal diffusion coefficients
in pure NO are too large by roughly a factor of 2 on average. By comparison,
the calculated admixture diffusion coefficients are much closer to
the measurements and exhibit many of their qualitative features. We
note that the two-term approximation is in error by up to 40\% for
the longitudinal diffusion coefficient measurements near 500 Td. It
follows that a multi-term Boltzmann solver is also necessary here
for the accurate calculation of characteristic energies. Fig. \ref{fig:Comparison-of-transport}(c)
shows ionisation coefficients calculated for our initial set that
are generally larger than those measured by Lakshminarasimha \textit{et
al}. \citep{Lakshminarasimha1977} (up to $75\%$ larger in some instances).
Quantitatively, there is also room for improvement here. Fig. \ref{fig:Comparison-of-transport}(d)
shows the near absence of attachment in our simulations below $\sim15\ \mathrm{Td}$,
in stark contrast to the measurements of Parkes \textit{et al}. \citep{Parkes1972}.
This indicates the presence of a low-energy attachment process that
is currently missing from our initial set. Fig. \ref{fig:Comparison-of-transport}(e)
shows a good quantitative agreement between our calculated transverse
characteristic energies and those measured by Lakshminarasimha \textit{et
al}. \citep{Lakshminarasimha1977}. However, the same cannot be said
for the measurements of Mechlińska-Drewko \textit{et al}. \citep{Mechlinska-Drewko1999}
which, while qualitatively similar, tend to be much smaller than what
we calculate. Fig. \ref{fig:Comparison-of-transport}(f) exhibits
a disagreement between our calculated longitudinal characteristic
energies and the pure NO measurements of Mechlińska-Drewko \textit{et
al}. \citep{Mechlinska-Drewko1999} and Takeuchi \textit{et al}. \citep{Takeuchi2001},
both of which are consistently smaller than what we calculate. The
agreement is much better for the mixture measurements of Takeuchi
\textit{et al}. \citep{Takeuchi2001}, although the simulations are
still larger than expected in the case of the 4.99\% NO in Ar measurements
below 3 Td.

\section{\label{sec:Cross-section-refinement}Cross section refinement using
data-driven swarm analysis}

In the previous section, we have seen clear discrepancies between
the transport coefficients calculated using our initial electron-NO
cross section set and those determined experimentally by a number
of authors. In this section, we apply machine learning in order to
refine our initial set and hopefully improve its agreement with the
measured swarm data plotted in Fig. \ref{fig:Comparison-of-transport}.

\subsection{Machine learning methodology}

To obtain a solution to the inverse swarm problem, we utilise the
artificial neural network of Stokes \textit{et al}. \citep{Stokes2019,Stokes2020,Stokes2021}:
\begin{equation}
\mathbf{y}\left(\mathbf{x}\right)=\left(\mathbf{A}_{4}\circ\mathrm{mish}\circ\mathbf{A}_{3}\circ\mathrm{mish}\circ\mathbf{A}_{2}\circ\mathrm{mish}\circ\mathbf{A}_{1}\right)\left(\mathbf{x}\right),\label{eq:neuralnet}
\end{equation}
where $\mathbf{A}_{n}\left(\mathbf{x}\right)\equiv\mathbf{W}_{n}\mathbf{x}+\mathbf{b}_{n}$
are affine mappings defined by dense \textit{weight} matrices $\mathbf{W}_{n}$
and \textit{bias} vectors $\mathbf{b}_{n}$, and $\mathrm{mish}\left(x\right)=x\tanh\left(\ln\left(1+e^{x}\right)\right)$
is a nonlinear \textit{activation function} \citep{Misra2019} that
is applied element-wise. The output vector, $\mathbf{y}$, contains
each NO cross section of interest as a function of energy, $\varepsilon$,
which is an element of the input vector, $\mathbf{x}$, alongside
all of the swarm measurements plotted in Fig. \ref{fig:Comparison-of-transport}.
It should be noted that we apply suitable logarithmic transformations
to ensure that all inputs and outputs of the network are dimensionless
and lie within $\left[-1,1\right]$. We specify that each bias vector
contains 256 parameters, with the exception of $\mathbf{b}_{4}$,
the size of which must match the number of cross sections in $\mathbf{y}$.
The weight matrices are all sized accordingly.

In order to train the neural network, Eq. \eqref{eq:neuralnet}, we
require an appropriate set of example solutions to the inverse swarm
problem. The choice of cross sections used for training is vital,
in order for the network to provide a self-consistent and physically-plausible
set of cross sections that are consistent within all the experimental
error bars (including those on the TCS). We derive such cross sections
from the LXCat project \citep{Pancheshnyi2012,Pitchford2017,Carbone2021},
as described in detail in Sec. \ref{subsec:Refined-cross-section}
for each of the cross sections considered. Once suitable training
cross sections are found, corresponding transport coefficients are
determined using our multi-term Boltzmann solver \citep{White2009,Boyle2017,White2018}
(here we use a four-term approximation). A small amount of random
noise is also multiplied by each transport coefficient to roughly
imitate random experimental error. The logarithm of this noise factor
is sampled from a normal distribution with a mean of 0 and standard
deviation of 0.03 for the drift velocities, 0.05 for the diffusion
coefficients, and 0.1 for the ionisation coefficients, attachment
coefficients and characteristic energies considered.

We implement and train the neural network, Eq. \eqref{eq:neuralnet},
using the \textit{Flux.jl} machine learning framework \citep{Innes2018}.
The network is initialised such that its biases are zero and its weights
are uniform random numbers, as described by Glorot and Bengio \citep{Glorot2010}.
Training is performed using the AdaBelief optimiser \citep{zhuang2020adabelief}
with Nesterov momentum \citep{Nesterov1983,Dozat2016}, step size
$\alpha=10^{-3}$, exponential decay rates $\beta_{1}=0.9$ and $\beta_{2}=0.999$,
and small parameter $\epsilon=10^{-8}$. At each iteration, the optimiser
is provided with a different batch of 4096 training examples, where
each batch consists of 32 training cross sections each evaluated at
128 random energies of the form $\varepsilon=10^{s}$ where $s\in\left[-2,3\right]$
is sampled from a continuous uniform distribution. For each batch,
the optimiser adjusts the neural network weights and biases with the
aim of further minimising the mean absolute error in solving the inverse
swarm problem for that batch. Training is continued for 150,000 iterations,
providing an equal number of potential solutions to the inverse swarm
problem. The quality of each of these solutions is subsequently assessed
by calculating their corresponding transport coefficients and comparing
these to the swarm data plotted in Fig. \ref{fig:Comparison-of-transport}.
The set that agrees best with the swarm measurements is presented
below as our ``refined'' set.

\subsection{\label{subsec:Refined-cross-section}Refined cross section set}

\begin{figure}
\begin{centering}
\includegraphics[scale=0.5]{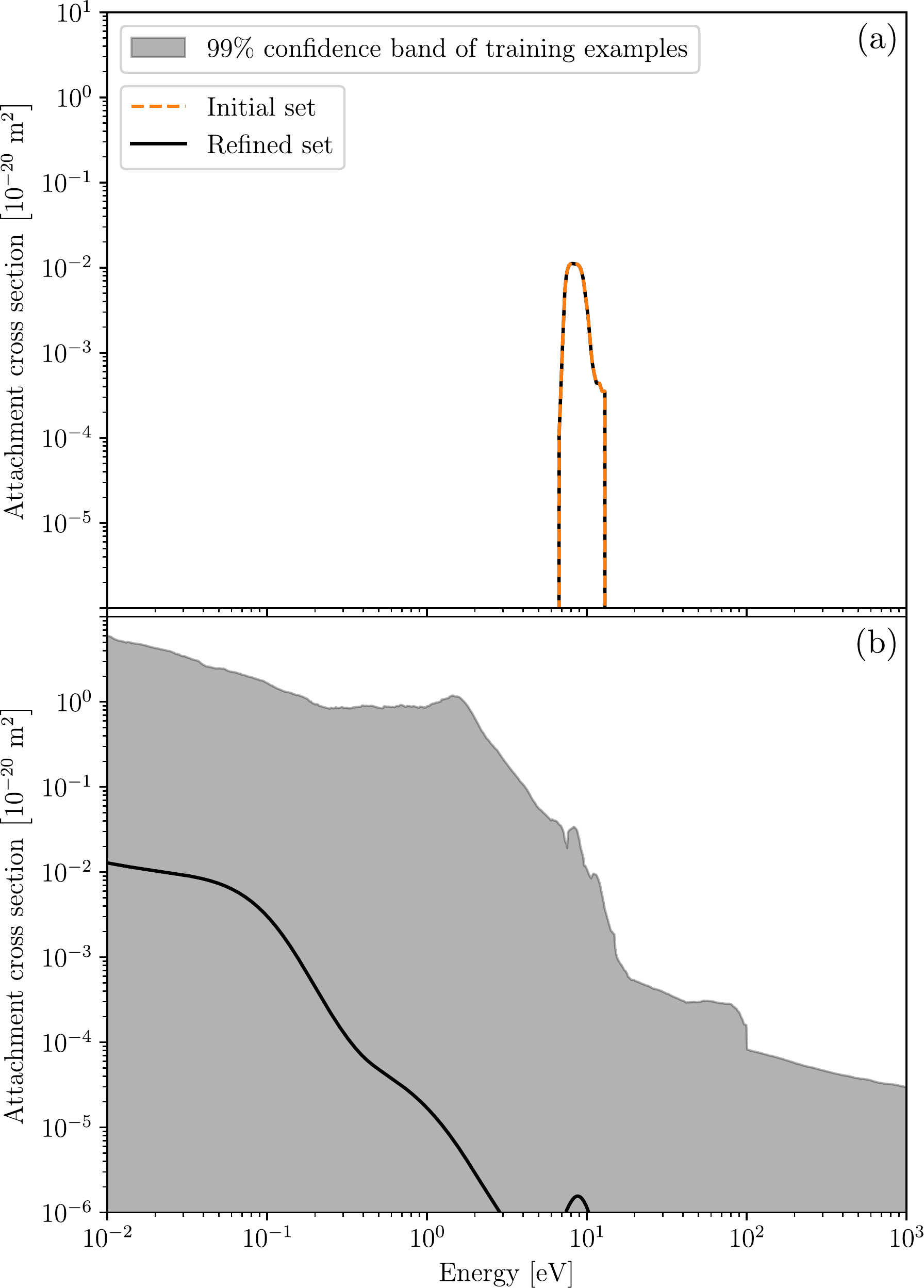}
\par\end{centering}
\caption{\label{fig:DEA-refined}Neural network regression result for an attachment
cross section, (b), present in the refined set alongside that from
the initial set, (a). See also legends in figure.}
\end{figure}
\begin{figure}
\begin{centering}
\includegraphics[scale=0.5]{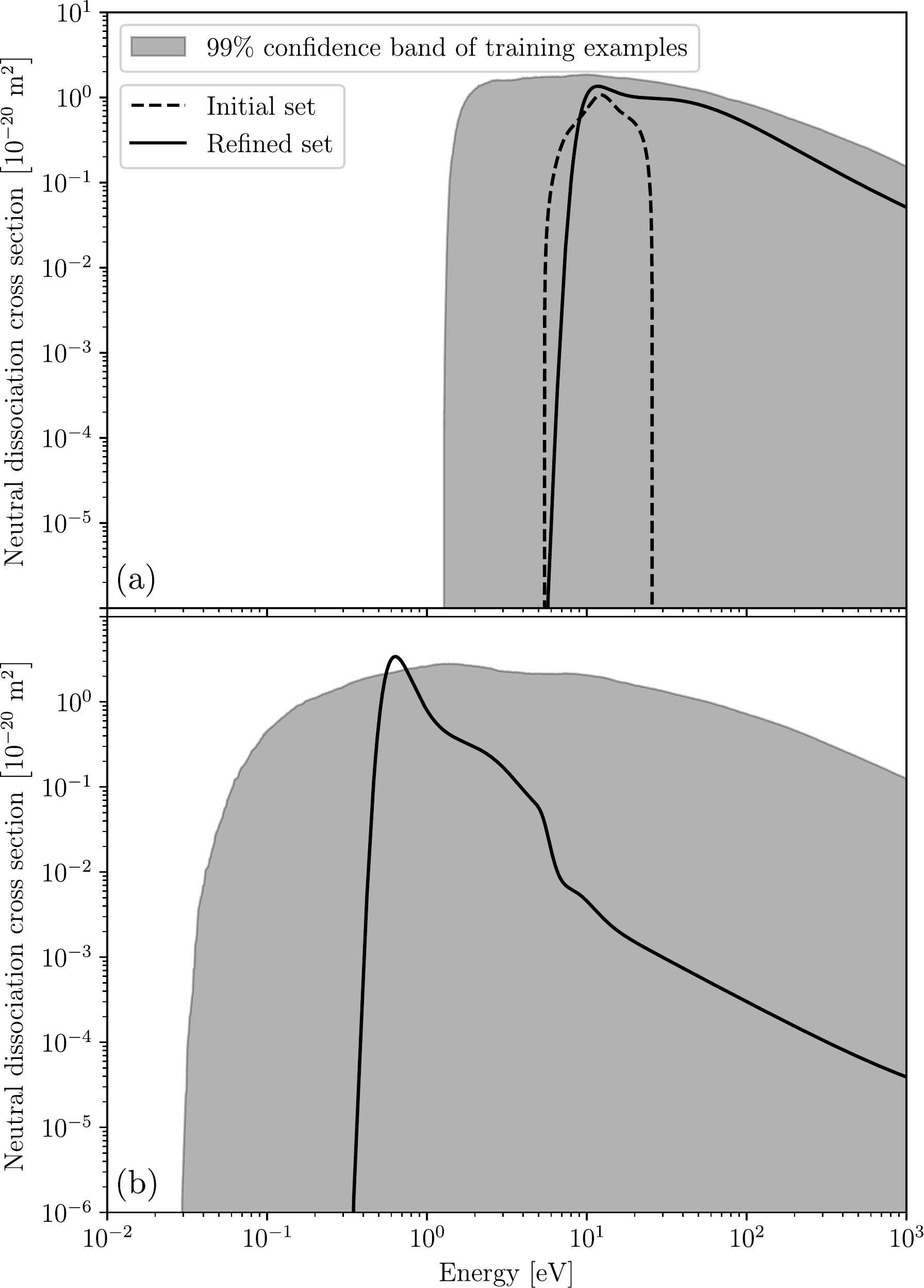}
\par\end{centering}
\caption{\label{fig:ND-refined}Neural network regression results for a pair
of NO neutral dissociation cross sections, one with a higher threshold
energy, (a), than the other, (b). The higher-threshold ND cross section
has a comparable threshold energy to the ND cross section from the
initial set, which is plotted alongside for comparison. See also legends
in figure.}
\end{figure}
\begin{figure}
\begin{centering}
\includegraphics[scale=0.5]{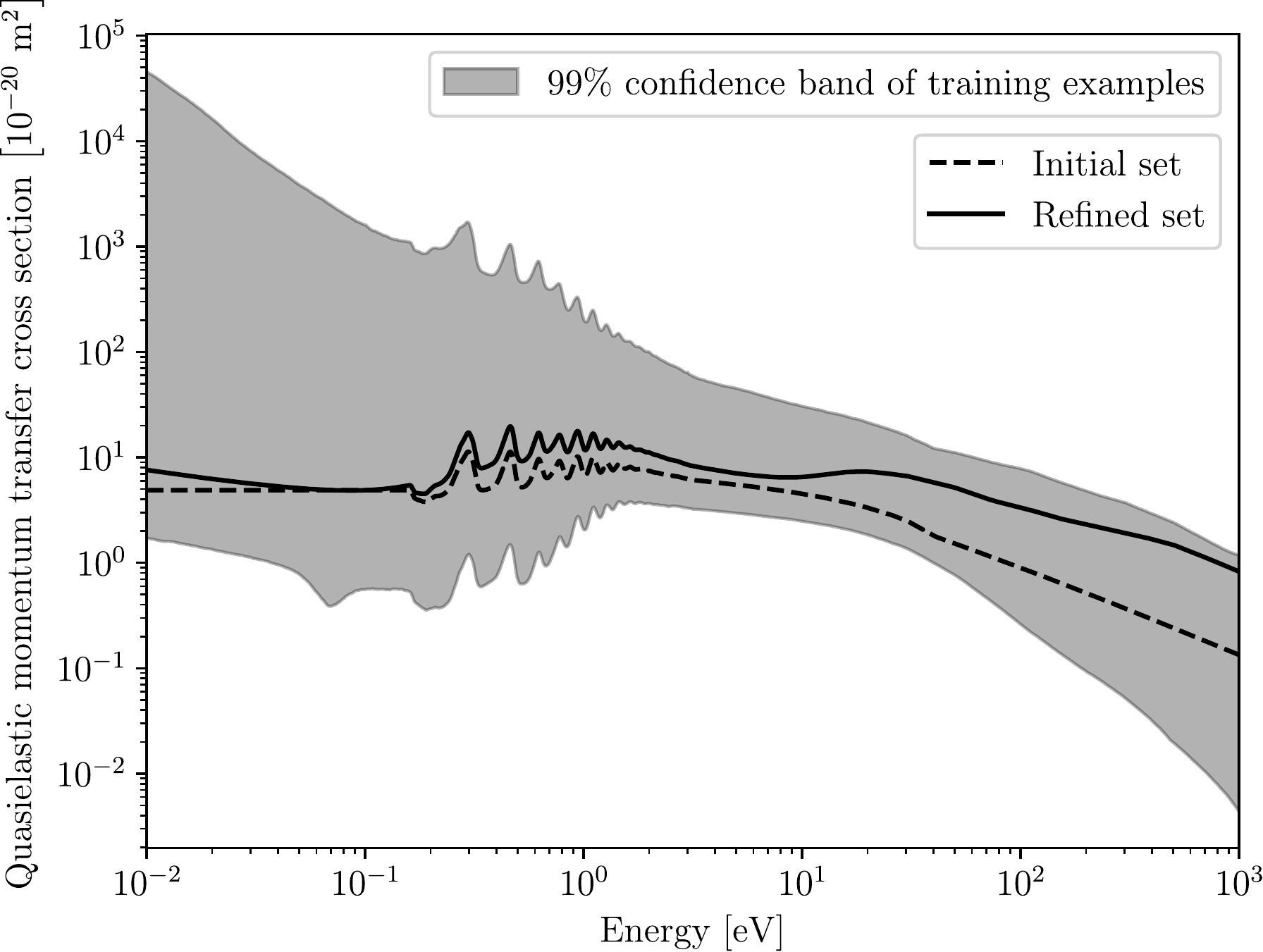}
\par\end{centering}
\caption{\label{fig:MTCS-refined}Neural network regression results for the
NO quasielastic MTCS, alongside the MTCS from the initial set for
comparison. See also legends in figure.}
\end{figure}
\begin{figure}
\begin{centering}
\includegraphics[scale=0.5]{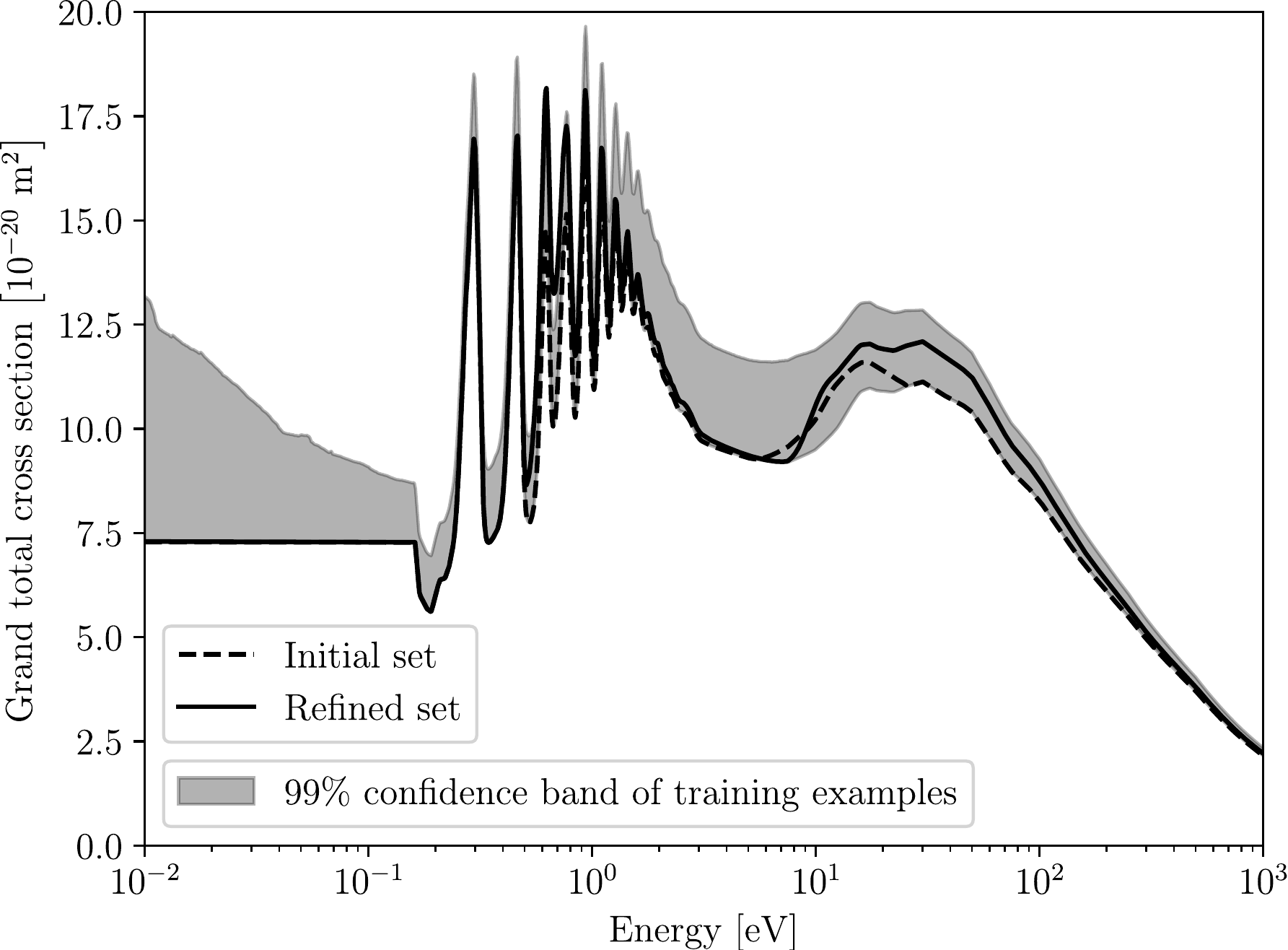}
\par\end{centering}
\caption{\label{fig:TCS-refined}Neural network regression results for the
NO TCS, alongside the initial set TCS for comparison. See also legends
in figure.}
\end{figure}
Using our initial cross section set as a base, we apply the neural
network, Eq. \eqref{eq:neuralnet}, to determine a selection of plausible
NO cross sections underlying the the swarm measurements shown in Fig.
\ref{fig:Comparison-of-transport}. In particular, we determine cross
sections for quasielastic (elastic+rotation) momentum transfer, neutral
dissociation, and dissociative electron attachment.

For dissociative electron attachment, we keep the cross section from
the initial set and determine an additional DEA cross section using
physically-plausible training data of the form:
\begin{equation}
\sigma\left(\varepsilon\right)=\sigma_{1}^{1-r}\left(\varepsilon\right)\sigma_{2}^{r}\left(\varepsilon\right),\label{eq:simplemixture}
\end{equation}
where $r\in\left[0,1\right]$ is a mixing ratio sampled from a continuous
uniform distribution, and $\sigma_{1}\left(\varepsilon\right)$ and
$\sigma_{2}\left(\varepsilon\right)$ are a random pair of attachment
cross sections from the LXCat project \citep{Pancheshnyi2012,Pitchford2017,Carbone2021}.
Beyond the implicit constraints due to the TCS (discussed below) and
the physical constraints inherent to the LXCat cross sections, we
otherwise leave the DEA training cross sections completely unconstrained.
The resulting confidence band of training examples is plotted in Fig.
\ref{fig:DEA-refined}(b), alongside the refined fit provided by the
neural network. Fig. \ref{fig:DEA-refined}(a), on the other hand,
simply shows that the DEA cross section from the initial set has remained
unchanged in the refined set. As indicated by the attachment coefficient
discrepancy shown in Fig. \ref{fig:Comparison-of-transport}(d), the
additional DEA cross section refined in Fig. \ref{fig:DEA-refined}(b)
is largest at low energies. Below 0.1 eV, the refined DEA is roughly
constant at $10^{-22}\ \mathrm{m^{2}}$, a magnitude comparable to
the peak of the other, higher-energy, DEA process kept from the initial
set. From 0.1 eV to 3 eV, the refined DEA cross section decays roughly
according to a power law before vanishing. The refinement increases
only very slightly in the vicinity of the other DEA process, near
9 eV, indicating that it is likely adequate as is.

For neutral dissociation, we found that fitting a single unconstrained
ND cross section resulted in a refinement with a threshold energy
much lower than that for the ND proposed in the initial set ($\sim5.5\ \mathrm{eV}$).
It was only after attempting a fit with two unconstrained ND processes
that a refined cross section of similar threshold energy arose. The
training data used for these two processes took the form:\textbf{
\begin{equation}
\sigma\left(\varepsilon\right)=\sigma_{1}^{1-r}\left(\varepsilon+\varepsilon_{1}-\varepsilon_{1}^{1-r}\varepsilon_{2}^{r}\right)\sigma_{2}^{r}\left(\varepsilon+\varepsilon_{2}-\varepsilon_{1}^{1-r}\varepsilon_{2}^{r}\right),\label{eq:mixture}
\end{equation}
}where $\sigma_{1}\left(\varepsilon\right)$ and $\sigma_{2}\left(\varepsilon\right)$
are a random pair of excitation cross sections from the LXCat project
\citep{Pancheshnyi2012,Pitchford2017,Carbone2021}, $\varepsilon_{1}$
and $\varepsilon_{2}$ are their respective threshold energies, and
$r\in\left[0,1\right]$ is a uniformly sampled mixing ratio. We additionally
use rejection sampling to enforce that the threshold energies of the
two ND processes are at least $1\ \mathrm{eV}$ apart. The resulting
ND training example confidence bands are shown in Fig. \ref{fig:ND-refined}.
Fig. \ref{fig:ND-refined}(a) indicates that the refined higher-threshold
ND process has a threshold energy that is almost identical to its
counterpart in the initial set (at $\sim5.6\ \mathrm{eV}$ versus
$\sim5.5\ \mathrm{eV}$). While this refinement increases slowly from
threshold compared to its counterpart, it ultimately reaches a slightly
larger peak magnitude ($1.35\times10^{-20}\ \mathrm{m}^{2}$ versus
$1.05\times10^{-20}\ \mathrm{m}^{2}$) at a slightly lower energy
(11.7 eV versus 12.3 eV). Beyond this turning point, the refinement
decays monotonically, transitioning to power-law above 50 eV before
falling to $5\times10^{-22}\ \mathrm{m^{2}}$ by 1000 eV. This is
in stark contrast to the initial ND cross section, which lacks a high-energy
tail due to its origin as a residual ICS from the TCS. Fig. \ref{fig:ND-refined}(b)
shows that the refined lower-threshold ND process has a threshold
energy of $\sim0.35\ \mathrm{eV}$. This refined cross section increases
rapidly from threshold, reaching a peak magnitude of $3.4\times10^{-20}\ \mathrm{m^{2}}$
at 0.63 eV. The subsequent decay here is also monotonic, albeit somewhat
more erratic, reaching a minimum of $3.9\times10^{-25}\ \mathrm{m^{2}}$
by 1000 eV.

For the quasielastic momentum transfer cross section, we take a slightly
different approach by using the neural network to determine a suitable
energy-dependent scaling of the elastic ICS, which upon application
would yield the refined quasielastic MTCS. The form of the scaling
factors that are used for training is:
\begin{equation}
\frac{\sigma\left(\varepsilon\right)}{\sigma_{\mathrm{ICS}}\left(\varepsilon\right)}=\left(\frac{\sigma_{\mathrm{QMTCS}}\left(\varepsilon\right)}{\sigma_{\mathrm{ICS}}\left(\varepsilon\right)}\right)^{1-r\left(\varepsilon\right)}\left(\frac{\sigma_{1}\left(\varepsilon\right)}{\sigma_{2}\left(\varepsilon\right)}\right)^{r\left(\varepsilon\right)},
\end{equation}
where $\sigma\left(\varepsilon\right)$ is the quasielastic MTCS used
for training, $\sigma_{\mathrm{ICS}}\left(\varepsilon\right)$ is
the present elastic ICS, $\sigma_{\mathrm{QMTCS}}\left(\varepsilon\right)$
is the present quasielastic MTCS (i.e. from the initial set), $\sigma_{1}\left(\varepsilon\right)$
and $\sigma_{2}\left(\varepsilon\right)$ are random elastic cross
sections formed from elastic cross sections on the LXCat project \citep{Pancheshnyi2012,Pitchford2017,Carbone2021}
that are combined using Eq. \eqref{eq:simplemixture}, and $r\left(\varepsilon\right)$
is a deterministic energy-dependent mixing ratio, used to constrain
the quasielastic MTCS training data to lie in the vicinity of the
experimental measurements of Mojarrabi \textit{et al.} \citep{Mojarrabi1995}:
\begin{equation}
r\left(\varepsilon\right)=\left\{ \begin{array}{cc}
1-0.4\frac{\ln\left(\frac{\varepsilon}{10^{-2}\ \mathrm{eV}}\right)}{\ln\left(\frac{1.5\ \mathrm{eV}}{10^{-2}\ \mathrm{eV}}\right)}, & 10^{-2}\ \mathrm{eV}\leq\varepsilon\leq1.5\ \mathrm{eV},\\
0.6, & 1.5\ \mathrm{eV}\leq\varepsilon\leq40\ \mathrm{eV},\\
0.6+0.4\frac{\ln\left(\frac{\varepsilon}{40\ \mathrm{eV}}\right)}{\ln\left(\frac{10^{3}\ \mathrm{eV}}{40\ \mathrm{eV}}\right)}, & 40\ \mathrm{eV}\leq\varepsilon\leq10^{3}\ \mathrm{eV}.
\end{array}\right.
\end{equation}
Further, to ensure each quasielastic MTCS decays at high energies,
we used rejection sampling to ensure that $\sigma_{1}\left(1000\ \mathrm{eV}\right)<\sigma_{2}\left(1000\ \mathrm{eV}\right)$.
The resulting quasielastic MTCS confidence band can be observed in
Fig. \ref{fig:MTCS-refined}, alongside the refined fit provided by
the neural network, and the initial quasielastic MTCS for comparison.
We see that the refined quasielastic MTCS is almost always larger
than that from our initial set, with the difference becoming greater
at higher energies. The greatest difference occurs at $10^{3}\ \mathrm{eV}$,
where the tail of the refined quasielastic MTCS is almost an order
of magnitude larger than that for the initial set.

When sampling ND and DEA cross sections for training, we further employ
rejection sampling to only keep cross section sets for training that
lie within $\pm7\%$ of the grand TCS of our initial set (while also
accounting for the uncertainty in the elastic ICS of $\pm25\%$).
The resulting confidence band in the grand TCS, due to the training
examples, is plotted in Fig. \ref{fig:TCS-refined}, alongside our
initial and refined grand TCSs. The refined TCS is roughly the same
as the initial TCS up until 10 eV, beyond which the refinement exceeds
that we initially proposed. The greatest difference arises at 26 eV,
with an increase in magnitude of 89\%.

\section{\label{sec:Multi-term-Boltzmann-equation-refined}Multi-term Boltzmann
equation analysis of our refined set}

\subsection{Consistency with swarm measurements}

\begin{figure}
\begin{centering}
\includegraphics[scale=0.5]{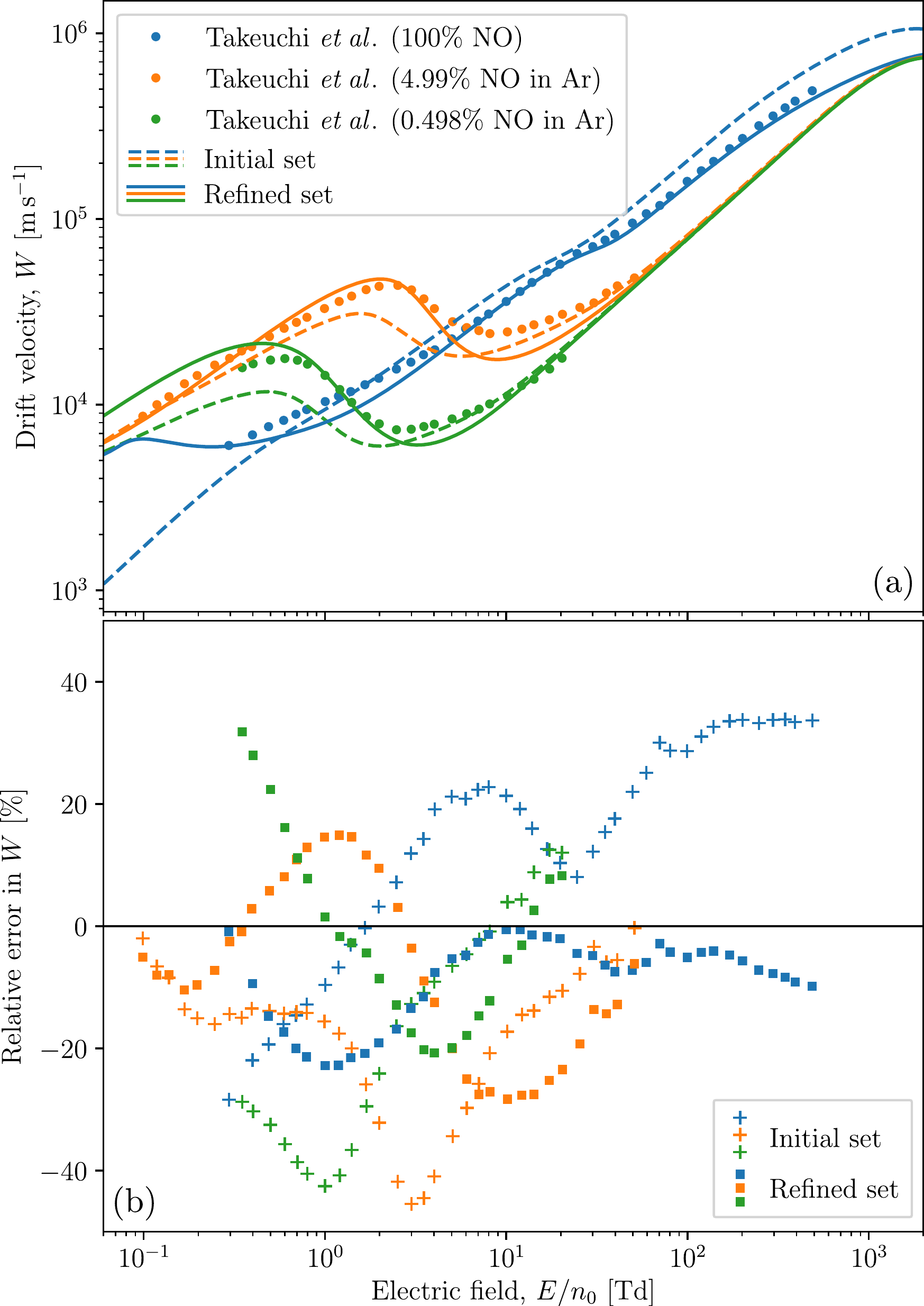}
\par\end{centering}
\caption{\label{fig:Drift-velocity}(a) Simulated arrival time spectra drift
velocities from both our initial data base and our refined data base,
compared to swarm measurements from the literature. (b) Corresponding
percentage errors in the simulated values relative to the swarm measurements.
See also legends in figures.}
\end{figure}
\begin{figure}
\begin{centering}
\includegraphics[scale=0.5]{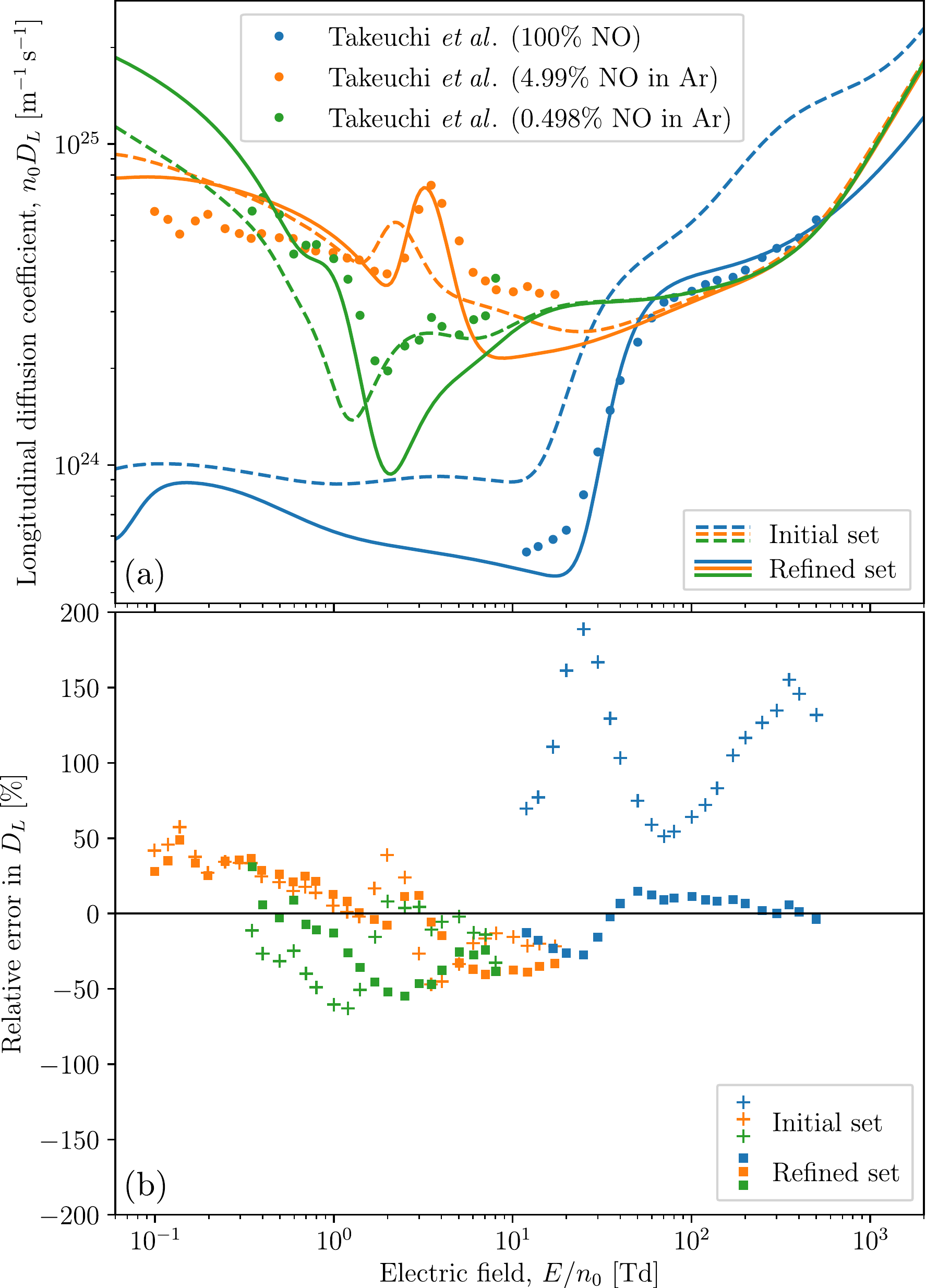}
\par\end{centering}
\caption{\label{fig:Diffusion-coefficient}(a) Simulated longitudinal diffusion
coefficients from both our initial data base and our refined data
base, compared to swarm measurements from the literature. (b) Corresponding
percentage errors in the simulated values relative to the swarm measurements.
See also legends in figures.}
\end{figure}
\begin{figure}
\begin{centering}
\includegraphics[scale=0.5]{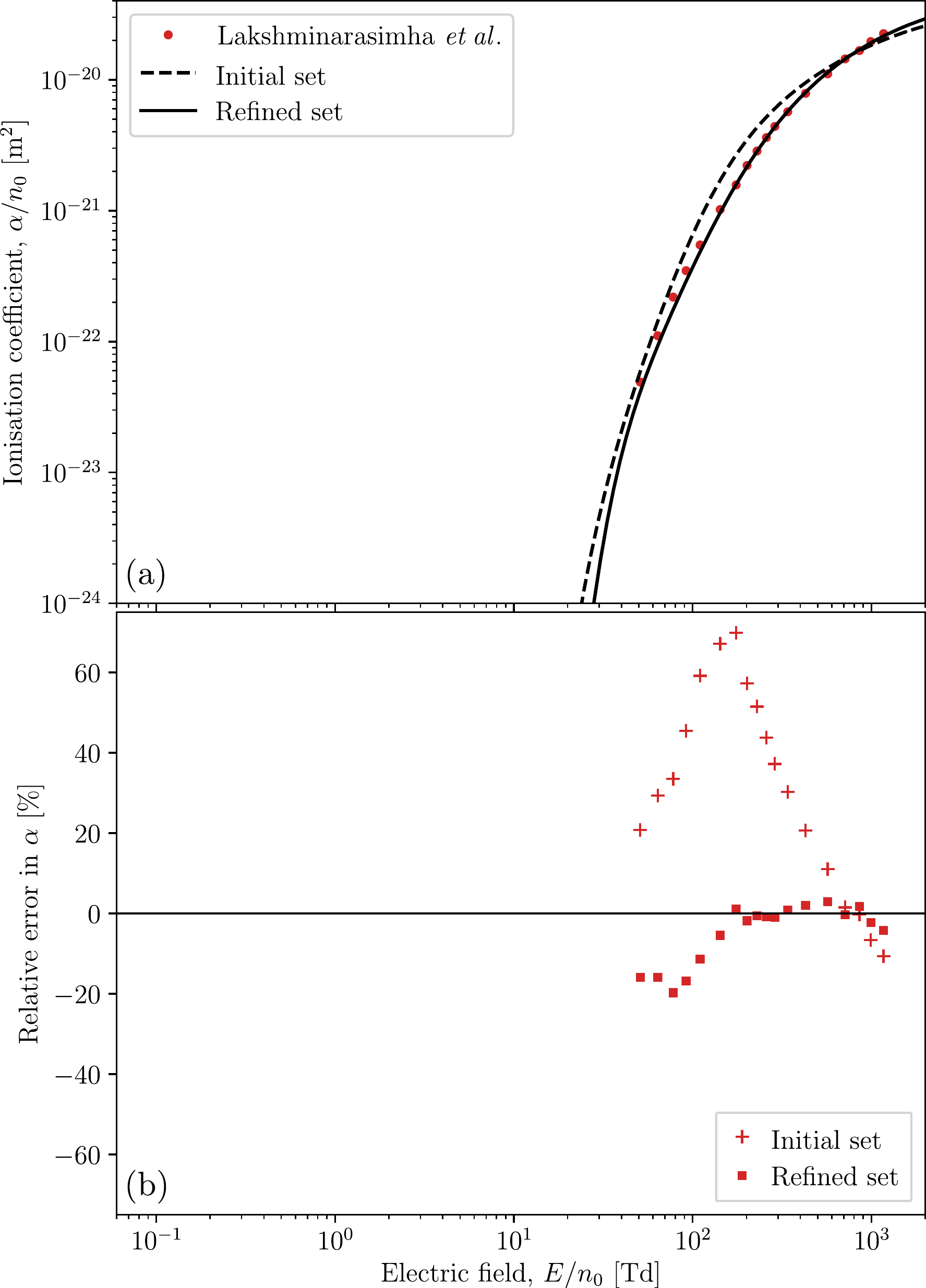}
\par\end{centering}
\caption{\label{fig:Ionisation-coefficient}(a) Simulated ionisation coefficients
from both our initial data base and our refined data base, compared
to swarm measurements from the literature. (b) Corresponding percentage
errors in the simulated values relative to the swarm measurements.
See also legends in figures.}
\end{figure}
\begin{figure}
\begin{centering}
\includegraphics[scale=0.5]{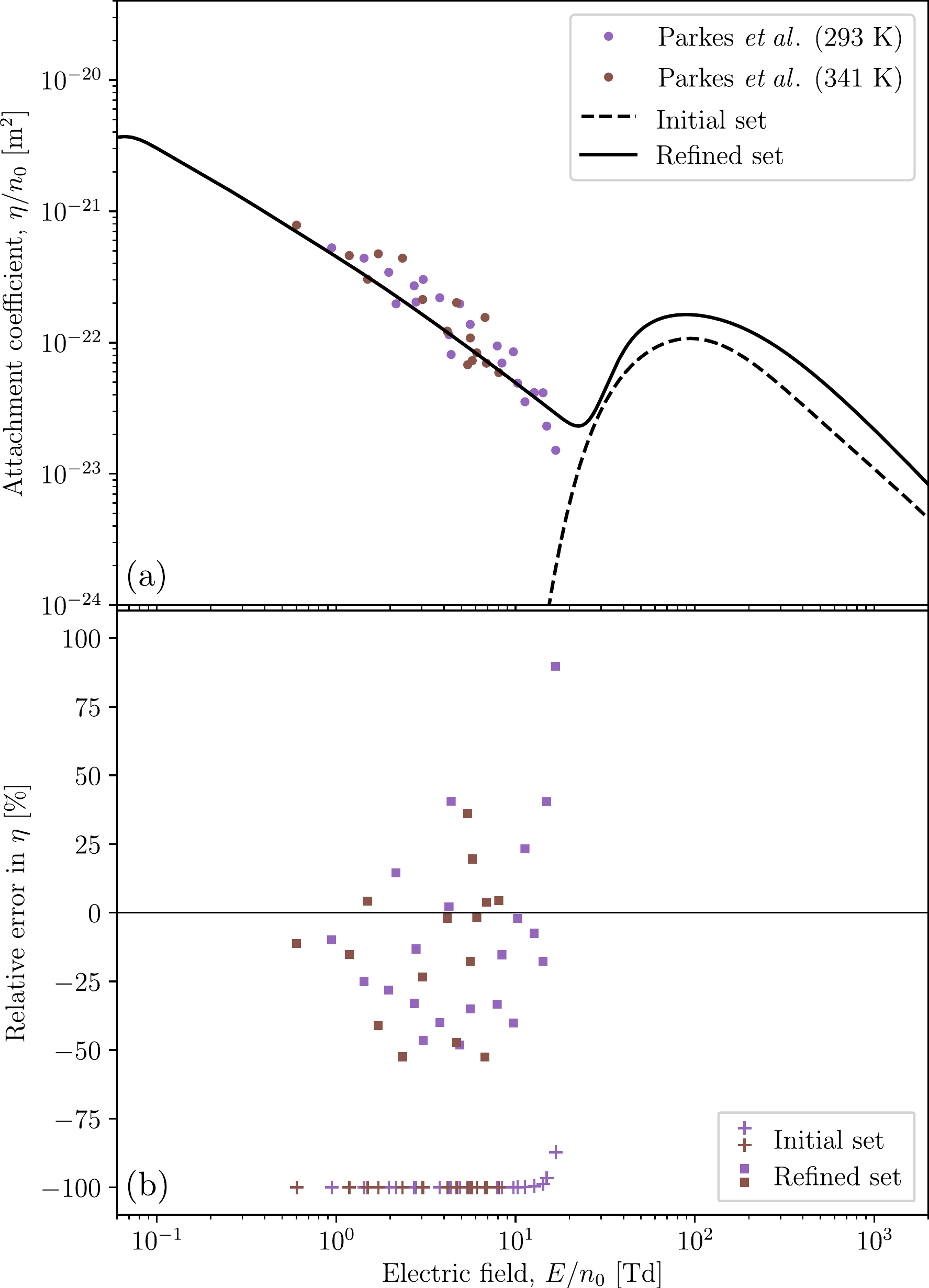}
\par\end{centering}
\caption{\label{fig:Attachment-coefficient}(a) Simulated attachment coefficients
from both our initial data base and our refined data base, compared
to swarm measurements from the literature. (b) Corresponding percentage
errors in the simulated values relative to the swarm measurements.
See also legends in figures.}
\end{figure}
\begin{figure}
\begin{centering}
\includegraphics[scale=0.5]{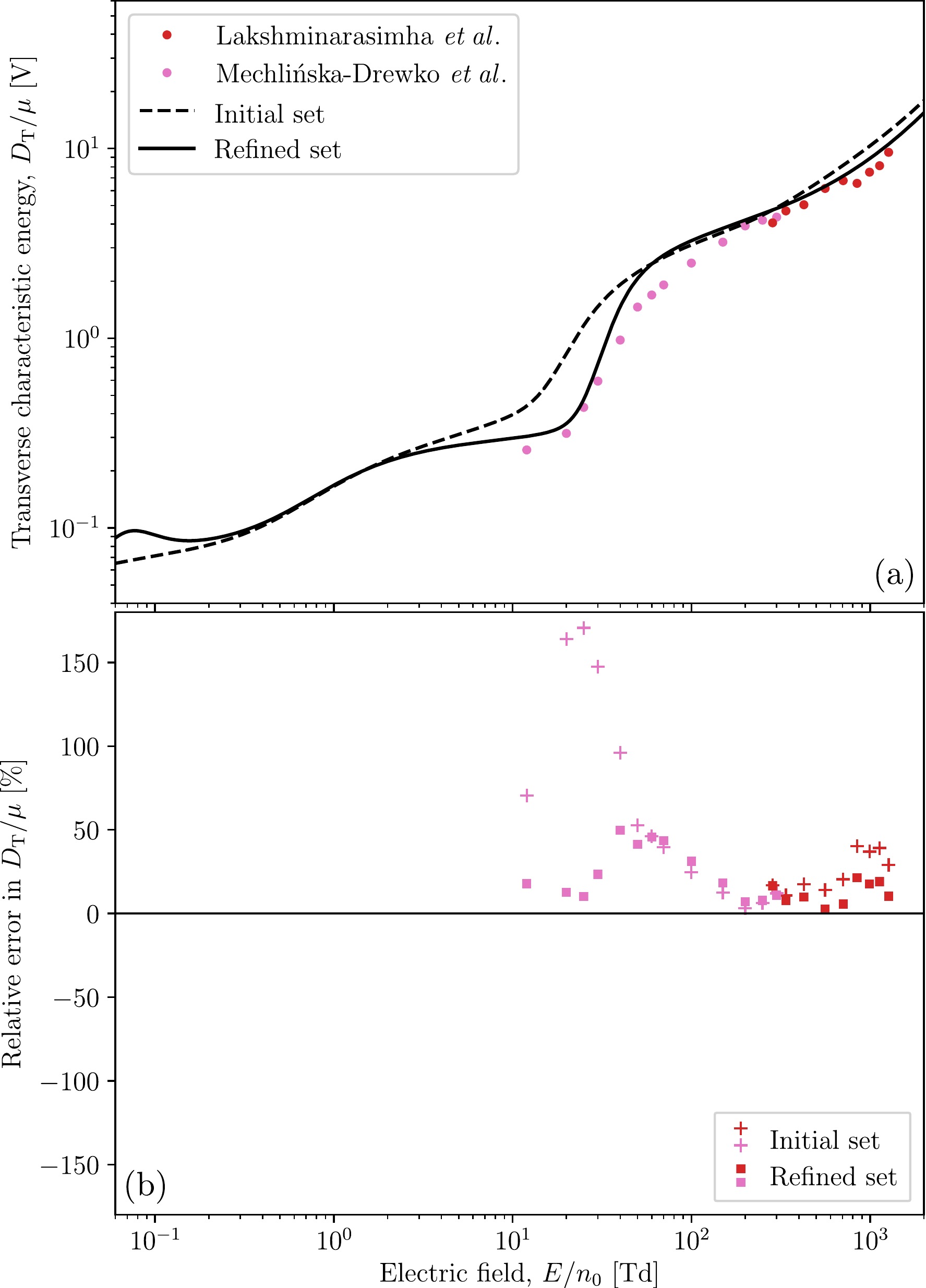}
\par\end{centering}
\caption{\label{fig:DTonmu}(a) Simulated transverse characteristic energies
from both our initial data base and our refined data base, compared
to swarm measurements from the literature. (b) Corresponding percentage
errors in the simulated values relative to the swarm measurements.
See also legends in figures.}
\end{figure}
\begin{figure}
\begin{centering}
\includegraphics[scale=0.5]{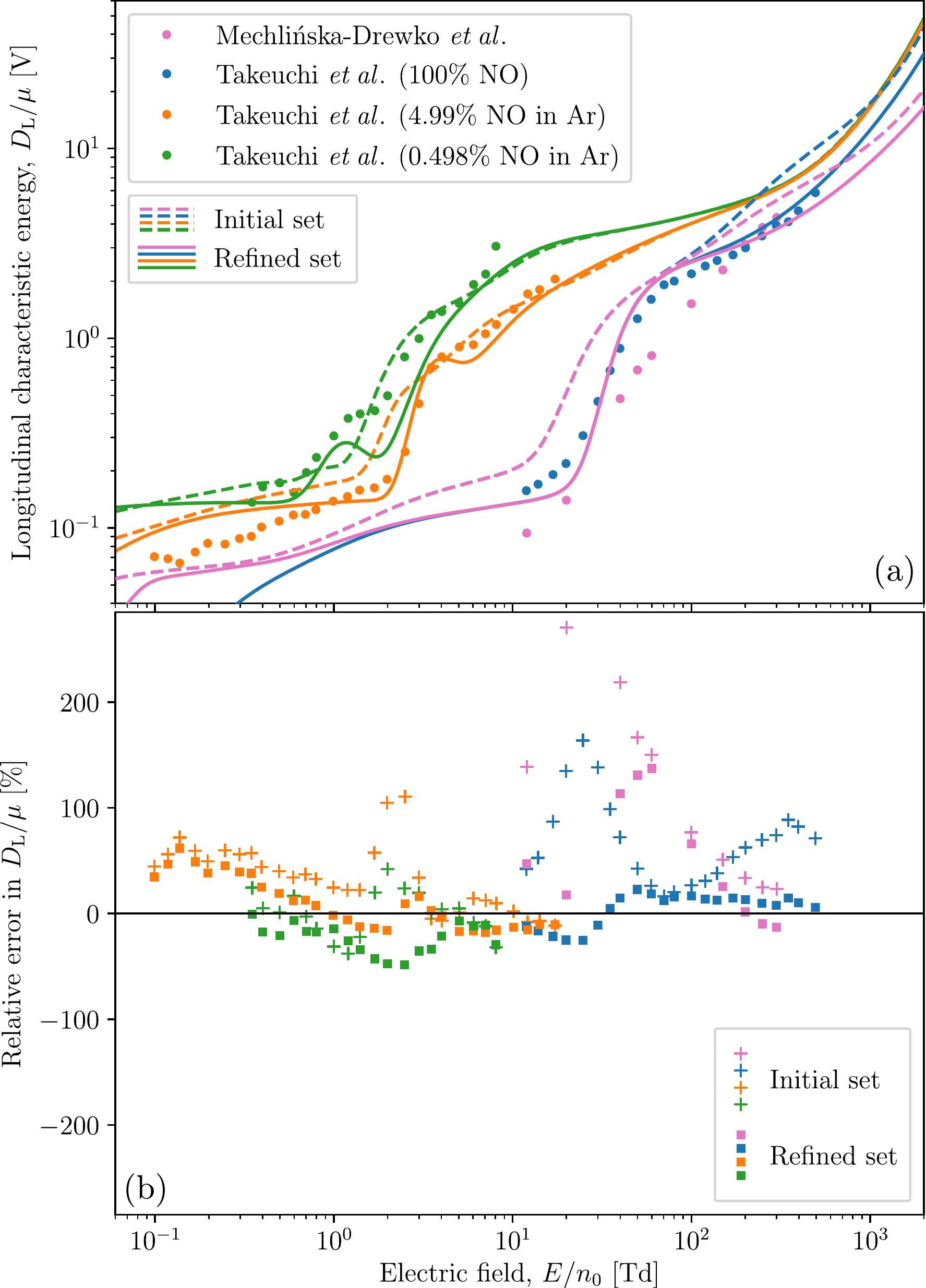}
\par\end{centering}
\caption{\label{fig:DLonmu}(a) Simulated longitudinal characteristic energies
from both our initial data base and our refined data base, compared
to swarm measurements from the literature. (b) Corresponding percentage
errors in the simulated values relative to the swarm measurements.
See also legends in figures.}
\end{figure}
Using our refined set of electron-NO cross sections, we plot revised
simulated transport coefficients in Figs. \ref{fig:Drift-velocity}(a)–\ref{fig:DLonmu}(a)
for comparison to the swarm measurements used to perform the refinement,
as well as to the transport coefficients calculated previously using
our initial set. These transport coefficients are accompanied by corresponding
percentage difference (error) plots in Figs. \ref{fig:Drift-velocity}(b)–\ref{fig:DLonmu}(b).
Here, positive percentage differences indicate that our simulated
transport coefficients exceed their experimentally-measured counterparts.
Fig. \ref{fig:Drift-velocity} shows that our refined set has brought
the simulated drift velocities much closer to the experimental measurements
of Takeuchi \textit{et al}. \citep{Takeuchi2001}. This agreement
is particularly good for the pure NO measurements above 3 Td, with
the discrepancies at lower $E/n_{0}$ possibly attributable to the
drift velocities in this regime being estimated from pressure-dependent
measurements \citep{Takeuchi2001}. Fig. \ref{fig:Diffusion-coefficient}
shows a large improvement after refinement for the pure NO diffusion
measurements of Takeuchi \textit{et al}. \citep{Takeuchi2001}. However,
the outcome is somewhat more mixed in the case of diffusion measurements
in the NO-Ar mixtures, with the refined set improving the agreement
with some of the admixture measurements of Takeuchi \textit{et al}.
\citep{Takeuchi2001}, while worsening it with others. Fig. \ref{fig:Ionisation-coefficient}
shows a significantly improved agreement between the ionisation coefficients
of the refined set and the measurements of Lakshminarasimha \textit{et
al}. \citep{Lakshminarasimha1977}. As this agreement was achieved
without requiring adjustments to the initial TICS, this lends further
credence to the overall self-consistency of the refinements that were
made. Fig. \ref{fig:Attachment-coefficient} indicates that the attachment
coefficient after refinement is non-zero below 20 Td, and in line
with the measurements of Parkes \textit{et al}. \citep{Parkes1972}.
The accuracy here is now generally within $\pm50\%$, which is of
the order of the uncertainty in the experimental measurements. Fig.
\ref{fig:DTonmu} shows a transverse characteristic energy after refinement
that agrees much better with the measurements of Mechlińska-Drewko
\textit{et al}. \citep{Mechlinska-Drewko1999} and Lakshminarasimha
\textit{et al}. \citep{Lakshminarasimha1977}, while Fig. \ref{fig:DLonmu}
shows similar improvements for the simulated longitudinal characteristic
energy as compared to the measurements of Mechlińska-Drewko \textit{et
al}. \citep{Mechlinska-Drewko1999} and Takeuchi \textit{et al}. \citep{Takeuchi2001}.
In this case, where the neural network was provided with both conflicting
sets of pure NO measurements, the refined fit is seen here to be more
consistent with the measurements of Takeuchi \textit{et al}. \citep{Takeuchi2001}
over those of Mechlińska-Drewko \textit{et al}. \citep{Mechlinska-Drewko1999}.

\subsection{Transport calculations for pure NO}

\begin{figure}
\begin{centering}
\includegraphics[scale=0.45]{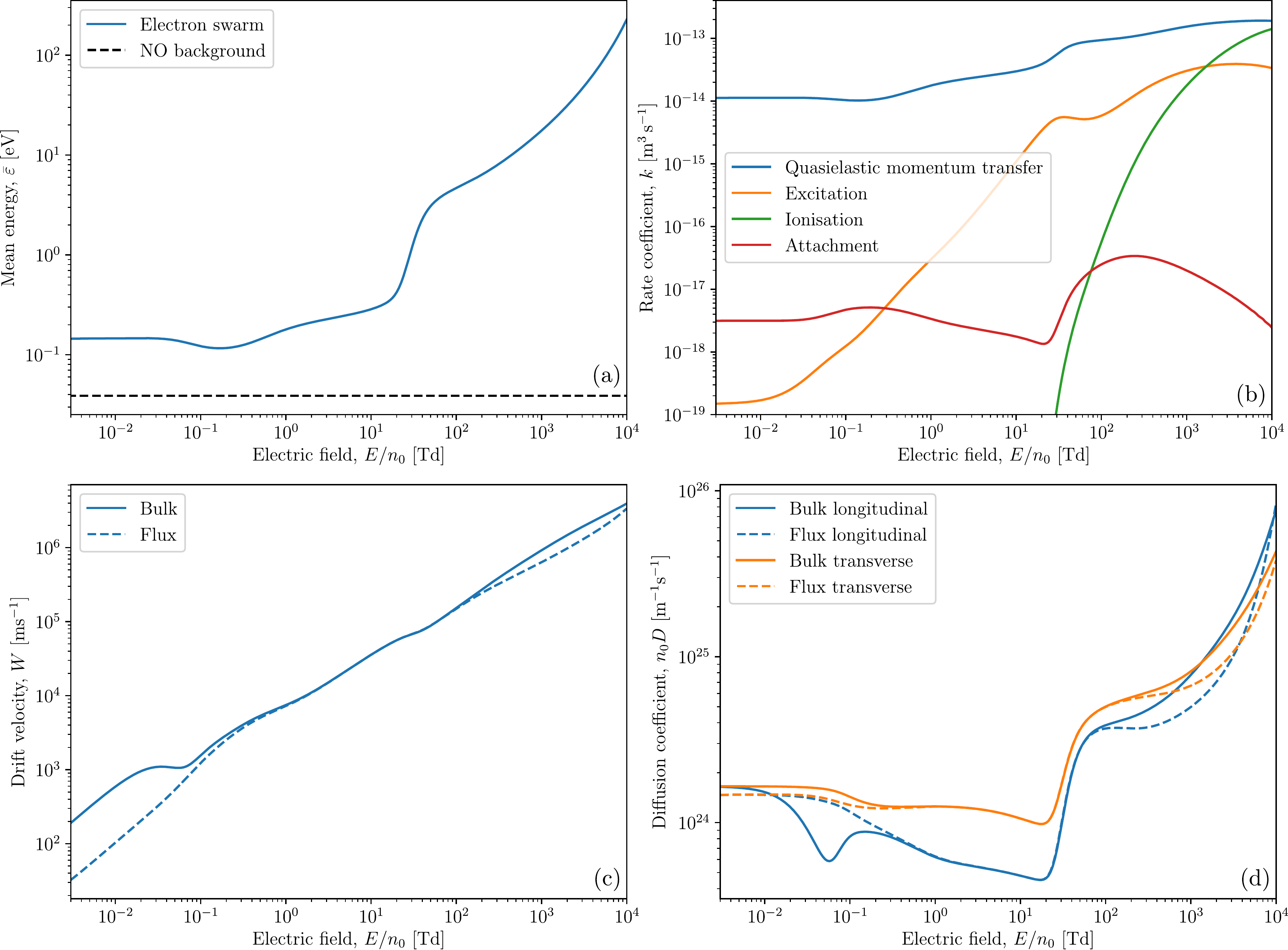}
\par\end{centering}
\caption{\label{fig:Calculated-mean-electron}Calculated mean electron energies,
(a), rate coefficients, (b), drift velocities, (c), and diffusion
coefficients, (d), for electrons in NO at $T=300\ \mathrm{K}$ (with
thermal energy $\frac{3}{2}k_{\mathrm{B}}T\approx38.8\ \mathrm{meV}$)
over a large range of reduced electric fields. See also the legends
for further details.}
\end{figure}
In Fig. \ref{fig:Calculated-mean-electron}, we apply our multi-term
Boltzmann solver to our refined cross section set, in order to determine
a variety of transport coefficients for electrons in NO at $T=300\ \mathrm{K}$
over a large range of reduced electric fields, $E/n_{0}$, varying
from $3\times10^{-3}\ \mathrm{Td}$ to $10^{4}\ \mathrm{Td}$. We
find that at least an eight-term approximation is necessary for all
the considered NO transport coefficients to be accurate to within
1\%. Fig. \ref{fig:Calculated-mean-electron}(a) shows a plot of mean
electron energy alongside that for the background NO. In the low-field
regime, the mean electron energy is $\sim146\ \mathrm{meV}$, which
is substantially higher than the thermal background of $\frac{3}{2}k_{\mathrm{B}}T\approx38.8\ \mathrm{meV}$,
a consequence of the attachment heating resulting from the refined
DEA cross section in Fig. \ref{fig:DEA-refined}(b). As $E/n_{0}$
is increased, the onset of the vibrational excitation channels contributes
to a net cooling of the electrons, causing the mean electron energy
to reach a minimum of $\sim116\ \mathrm{meV}$ at $\sim0.17\ \mathrm{Td}$.
Beyond that point, increasing $E/n_{0}$ further causes the mean electron
energy to increase monotonically due to heating by the field, with
its ascent occasionally slowing due to the onset of additional excitation
channels ($\sim1\ \mathrm{Td}$) and attachment and ionisation processes
($\sim200\ \mathrm{Td}$). Fig. \ref{fig:Calculated-mean-electron}(b)
shows the rate coefficients for quasielastic momentum transfer, summed
discrete excitation, ionisation and attachment. The quasielastic momentum
transfer rate coefficient is highly correlated with the mean electron
energy, although this correlation tapers off near $10^{4}\ \mathrm{Td}$
with a decrease with increasing $E/n_{0}$. Summed excitation and
ionisation rate coefficients both increase as expected with increasing
$E/n_{0}$. The summed attachment coefficient also behaves as expected,
with the low-energy and high-energy attachment cross sections resulting
in separate maxima at $\sim0.19\ \mathrm{Td}$ and $\sim250\ \mathrm{Td}$,
respectively. Fig. \ref{fig:Calculated-mean-electron}(c) shows the
bulk (centre-of-mass) and flux drift velocities of the swarm, which
are expected to differ when nonconservative processes are acting.
For example, in the low-field regime, the preferential attachment
of lower-energy electrons towards the back of the swarm results in
a forward shift in the centre of mass, yielding a bulk drift velocity
that is larger compared to its flux counterpart. Here, the bulk drift
velocity also exhibits NDC between $\sim3.5\times10^{-2}\ \mathrm{Td}$
and $\sim5.5\times10^{-2}\ \mathrm{Td}$. At intermediate fields,
between $\sim1\ \mathrm{Td}$ and $\sim100\ \mathrm{Td}$, net nonconservative
effects are sufficiently small such that the bulk and flux drift velocities
coincide. Above $\sim100\ \mathrm{Td}$, the bulk drift velocity again
exceeds the flux, this time due to ionisation preferentially creating
electrons at the front of the swarm. Fig. \ref{fig:Calculated-mean-electron}(d)
shows the bulk and flux\textbf{ }diffusion coefficients in directions
both longitudinal and transverse to the field. At very low $E/n_{0}$,
when the electron energy distribution function (EEDF) is nearly Maxwellian,
the preferential attachment of low-energy electrons removes electrons
from the centre of the swarm over those from the edges, causing an
effective increase in the bulk diffusion coefficients compared to
their flux counterparts. This remains the case transversely as $E/n_{0}$
increases due to the symmetry of the swarm in the transverse direction.
Longitudinally, however, the additional power input by the field causes
faster, higher energy, electrons to congregate at the front of the
swarm, thus introducing an asymmetry in the longitudinal direction.
The subsequent attachment of lower-energy electrons at the back of
the swarm causes the bulk longitudinal diffusion coefficient to drop
below its flux counterpart from $\sim1.2\times10^{-2}\ \mathrm{Td}$,
onward. As with the drift velocities, for intermediate $E/n_{0}$
between roughly 1 Td and 100 Td, nonconservative effects are minimal
and the bulk and flux diffusion coefficients coincide.\textbf{ }At
higher $E/n_{0}$, above 100 Td, there is a significant increase in
bulk diffusion compared to flux in both the transverse and longitudinal
directions, which we attribute to the preferential\textbf{ }production
of electrons due to ionisation at the front and sides of the swarm. 

\section{\label{sec:Conclusion}Conclusion}

We have formed a comprehensive and self-consistent set of electron-NO
cross sections, by constructing an initial set from the literature
and then refining it using a multi-term Boltzmann equation analysis
of the available swarm transport data. This refinement was performed
automatically and objectively using a neural network model, Eq. \eqref{eq:neuralnet},
that was trained on cross sections derived from the LXCat project
\citep{Pancheshnyi2012,Pitchford2017,Carbone2021}, to ensure that
the refinements made were physically plausible. In summary, by using
our initial set as a base, we obtained from the swarm data a quasielastic
MTCS, a DEA cross section and a ND cross section. Compared to our
initial set, we confirmed that our resulting refined cross section
database was more consistent with the swarm data from which it was
derived. Lastly, we used our refined set to calculate a variety of
transport coefficients for electrons in NO across a large range of
reduced electric fields. Notably, this revealed significant heating
of the swarm above the thermal background due to our refined low-energy
attachment process.

The machine learning methodology we have employed in this work has
previously \citep{Stokes2020} been shown to produce cross section
sets that are comparable to those refined using conventional swarm
analysis, i.e. through manual iterative refinement by an expert. We
thus believe that our refined set of electron-NO cross sections is
of a similarly high quality. That said, we do acknowledge there is
still some room for improvement in the present fit and it is thus
fortunate that this machine learning approach makes it straightforward
to revisit NO as new swarm data, cross section constraints, or LXCat
training data becomes available. On this note, we highlight that there
are presently no swarm measurements available for the \textit{effective}
first Townsend ionisation coefficient. We believe performing such
measurements would be a worthwhile future endeavour, as they would
allow one to quantify the attachment coefficient in the electropositive
regime (above $\sim75\ \mathrm{Td}$) and thus allow for the possibility
of refining the DEA cross section above $\sim4\ \mathrm{eV}$.

Given the ill-posed nature of the inverse swarm problem, we also acknowledge
that our refined set carries with it some uncertainty. In this sense,
an artificial neural network of the form of Eq. \eqref{eq:neuralnet}
is not ideal as it does not provide uncertainty quantification. In
the future, we may address this by using an appropriate alternative
machine learning model \citep{Bishop1994,Sohn2015,Mirza2014,Dinh2014,Dinh2016,Kingma2018}.
We also plan to apply our data-driven swarm analysis to determine
complete and self-consistent cross section sets for other molecules
of biological interest, including water \citep{White2014}.

\ack{}{}

The authors gratefully acknowledge the financial support of the Australian
Research Council through the Discovery Projects Scheme (Grant \#DP180101655).

\section*{Data Availability Statement}

The data that supports the findings of this study are available within
the article.

\appendix

\bibliographystyle{iopart-num}
\addcontentsline{toc}{section}{\refname}\bibliography{references}

\providecommand{\newblock}{}
\begin{thebibliography}{10}
\expandafter\ifx\csname url\endcsname\relax
  \def\url#1{{\tt #1}}\fi
\expandafter\ifx\csname urlprefix\endcsname\relax\def\urlprefix{URL }\fi
\providecommand{\eprint}[2][]{\url{#2}}

\bibitem{Kong2009}
Kong M~G, Kroesen G, Morfill G, Nosenko T, Shimizu T, {Van Dijk} J and
  Zimmermann J~L 2009 {\em New Journal of Physics\/} {\bf 11} 115012 ISSN
  13672630 \urlprefix\url{https://doi.org/10.1088/1367-2630/11/11/115012}

\bibitem{Samukawa2012}
Samukawa S, Hori M, Rauf S, Tachibana K, Bruggeman P, Kroesen G, Whitehead J~C,
  Murphy A~B, Gutsol A~F, Starikovskaia S, Kortshagen U, Boeuf J~P, Sommerer
  T~J, Kushner M~J, Czarnetzki U and Mason N 2012 {\em Journal of Physics D:
  Applied Physics\/} {\bf 45} 253001 ISSN 0022-3727
  \urlprefix\url{https://doi.org/10.1088/0022-3727/45/25/253001}

\bibitem{Bruggeman2016}
Bruggeman P~J, Kushner M~J, Locke B~R, Gardeniers J~G, Graham W~G, Graves D~B,
  Hofman-Caris R~C, Maric D, Reid J~P, Ceriani E, {Fernandez Rivas} D, Foster
  J~E, Garrick S~C, Gorbanev Y, Hamaguchi S, Iza F, Jablonowski H, Klimova E,
  Kolb J, Krcma F, Lukes P, MacHala Z, Marinov I, Mariotti D, {Mededovic
  Thagard} S, Minakata D, Neyts E~C, Pawlat J, Petrovic Z~L, Pflieger R, Reuter
  S, Schram D~C, Schr{\"{o}}ter S, Shiraiwa M, Tarabov{\'{a}} B, Tsai P~A,
  Verlet J~R, {Von Woedtke} T, Wilson K~R, Yasui K and Zvereva G 2016 {\em
  Plasma Sources Science and Technology\/} {\bf 25} 053002 ISSN 13616595
  \urlprefix\url{https://doi.org/10.1088/0963-0252/25/5/053002}

\bibitem{Adamovich2017}
Adamovich I, Baalrud S~D, Bogaerts A, Bruggeman P~J, Cappelli M, Colombo V,
  Czarnetzki U, Ebert U, Eden J~G, Favia P, Graves D~B, Hamaguchi S, Hieftje G,
  Hori M, Kaganovich I~D, Kortshagen U, Kushner M~J, Mason N~J, Mazouffre S,
  Thagard S~M, Metelmann H~R, Mizuno A, Moreau E, Murphy A~B, Niemira B~A,
  Oehrlein G~S, Petrovic Z~L, Pitchford L~C, Pu Y~K, Rauf S, Sakai O, Samukawa
  S, Starikovskaia S, Tennyson J, Terashima K, Turner M~M, {Van De Sanden} M~C
  and Vardelle A 2017 {\em Journal of Physics D: Applied Physics\/} {\bf 50}
  323001 ISSN 13616463 \urlprefix\url{https://doi.org/10.1088/1361-6463/aa76f5}

\bibitem{Shekhter2005}
Shekhter A~B, Serezhenkov V~A, Rudenko T~G, Pekshev A~V and Vanin A~F 2005 {\em
  Nitric Oxide\/} {\bf 12} 210--219 ISSN 10898603
  \urlprefix\url{https://doi.org/10.1016/j.niox.2005.03.004}

\bibitem{Kim2016}
Kim S~J and Chung T~H 2016 {\em Scientific Reports\/} {\bf 6} 20332 ISSN
  2045-2322 \urlprefix\url{https://doi.org/10.1038/srep20332}

\bibitem{Li2017}
Li Y, {Ho Kang} M, {Sup Uhm} H, {Joon Lee} G, {Ha Choi} E and Han I 2017 {\em
  Scientific Reports\/} {\bf 7} 45781 ISSN 2045-2322
  \urlprefix\url{https://doi.org/10.1038/srep45781}

\bibitem{Tanaka2016}
Tanaka H, Brunger M~J, Campbell L, Kato H, Hoshino M and Rau A~R 2016 {\em
  Reviews of Modern Physics\/} {\bf 88} 025004 ISSN 15390756
  \urlprefix\url{https://doi.org/10.1103/RevModPhys.88.025004}

\bibitem{Mayer1921}
Mayer H~F 1921 {\em Annalen der Physik\/} {\bf 369} 451--480 ISSN 00033804
  \urlprefix\url{https://doi.org/10.1002/andp.19213690503}

\bibitem{Ramsauer1921}
Ramsauer C 1921 {\em Annalen der Physik\/} {\bf 369} 513--540 ISSN 15213889
  \urlprefix\url{https://doi.org/10.1002/andp.19213690603}

\bibitem{Townsend1922}
Townsend J and Bailey V 1922 {\em The London, Edinburgh, and Dublin
  Philosophical Magazine and Journal of Science\/} {\bf 43} 593--600 ISSN
  1941-5982 \urlprefix\url{https://doi.org/10.1080/14786442208633916}

\bibitem{Frost1962}
Frost L~S and Phelps A~V 1962 {\em Physical Review\/} {\bf 127} 1621--1633 ISSN
  0031899X \urlprefix\url{https://doi.org/10.1103/PhysRev.127.1621}

\bibitem{Engelhardt1963}
Engelhardt A~G and Phelps A~V 1963 {\em Physical Review\/} {\bf 131} 2115--2128
  ISSN 0031899X \urlprefix\url{https://doi.org/10.1103/PhysRev.131.2115}

\bibitem{Engelhardt1964}
Engelhardt A~G, Phelps A~V and Risk C~G 1964 {\em Physical Review\/} {\bf 135}
  A1566--A1574 ISSN 0031-899X
  \urlprefix\url{https://doi.org/10.1103/PhysRev.135.A1566}

\bibitem{Hake1967}
Hake R~D and Phelps A~V 1967 {\em Physical Review\/} {\bf 158} 70--84 ISSN
  0031899X \urlprefix\url{https://doi.org/10.1103/PhysRev.158.70}

\bibitem{Phelps1968}
Phelps A~V 1968 {\em Reviews of Modern Physics\/} {\bf 40} 399--410 ISSN
  00346861 \urlprefix\url{https://doi.org/10.1103/RevModPhys.40.399}

\bibitem{Stokes2019}
Stokes P~W, Cocks D~G, Brunger M~J and White R~D 2020 {\em Plasma Sources
  Science and Technology\/} {\bf 29} 055009 ISSN 13616595 (\textit{Preprint}
  \eprint{1912.05842}) \urlprefix\url{https://doi.org/10.1088/1361-6595/ab85b6}

\bibitem{Stokes2020}
Stokes P~W, Casey M~J, Cocks D~G, de~Urquijo J, Garc{\'{i}}a G, Brunger M~J and
  White R~D 2020 {\em Plasma Sources Science and Technology\/} {\bf 29} ISSN
  13616595 (\textit{Preprint} \eprint{2007.02762})
  \urlprefix\url{https://doi.org/10.1088/1361-6595/abb4f6}

\bibitem{Stokes2021}
Stokes P~W, Foster S~P, Casey M~J, Cocks D~G, Gonz{\'{a}}lez-Maga{\~{n}}a O,
  {De Urquijo} J, Garc{\'{i}}a G, Brunger M~J and White R~D 2021 {\em Journal
  of Chemical Physics\/} {\bf 154} 084306 ISSN 10897690
  \urlprefix\url{https://doi.org/10.1063/5.0043759}

\bibitem{DeUrquijo2019a}
{De Urquijo} J, Casey M~J, Serkovic-Loli L~N, Cocks D~G, Boyle G~J, Jones D~B,
  Brunger M~J and White R~D 2019 {\em Journal of Chemical Physics\/} {\bf 151}
  054309 ISSN 00219606 \urlprefix\url{https://doi.org/10.1063/1.5108619}

\bibitem{Morgan1991}
Morgan W~L 1991 {\em IEEE Transactions on Plasma Science\/} {\bf 19} 250--255
  ISSN 19399375 \urlprefix\url{https://doi.org/10.1109/27.106821}

\bibitem{Ceriotti2021}
Ceriotti M, Clementi C and {Anatole von Lilienfeld} O 2021 {\em The Journal of
  Chemical Physics\/} {\bf 154} 160401 ISSN 0021-9606
  \urlprefix\url{https://doi.org/10.1063/5.0051418}

\bibitem{Pancheshnyi2012}
Pancheshnyi S, Biagi S, Bordage M~C, Hagelaar G~J, Morgan W~L, Phelps A~V and
  Pitchford L~C 2012 {\em Chemical Physics\/} {\bf 398} 148--153 ISSN 03010104
  \urlprefix\url{https://doi.org/10.1016/j.chemphys.2011.04.020}

\bibitem{Pitchford2017}
Pitchford L~C, Alves L~L, Bartschat K, Biagi S~F, Bordage M~C, Bray I, Brion
  C~E, Brunger M~J, Campbell L, Chachereau A, Chaudhury B, Christophorou L~G,
  Carbone E, Dyatko N~A, Franck C~M, Fursa D~V, Gangwar R~K, Guerra V,
  Haefliger P, Hagelaar G~J, Hoesl A, Itikawa Y, Kochetov I~V, McEachran R~P,
  Morgan W~L, Napartovich A~P, Puech V, Rabie M, Sharma L, Srivastava R,
  Stauffer A~D, Tennyson J, de~Urquijo J, van Dijk J, Viehland L~A, Zammit M~C,
  Zatsarinny O and Pancheshnyi S 2017 {\em Plasma Processes and Polymers\/}
  {\bf 14} 1600098 ISSN 16128869
  \urlprefix\url{https://doi.org/10.1002/ppap.201600098}

\bibitem{Carbone2021}
Carbone E, Graef W, Hagelaar G, Boer D, Hopkins M~M, Stephens J~C, Yee B~T,
  Pancheshnyi S, van Dijk J and Pitchford L 2021 {\em Atoms\/} {\bf 9} 16 ISSN
  2218-2004
  \urlprefix\url{https://doi.org/https://doi.org/10.3390/atoms9010016}

\bibitem{Brunger2002}
Brunger M~J and Buckman S~J 2002 {\em Physics Reports\/} {\bf 357} 215--458
  ISSN 03701573 \urlprefix\url{https://doi.org/10.1016/S0370-1573(01)00032-1}

\bibitem{Itikawa2016}
Itikawa Y 2016 {\em Journal of Physical and Chemical Reference Data\/} {\bf 45}
  033106 ISSN 0047-2689 \urlprefix\url{https://doi.org/10.1063/1.4961372}

\bibitem{Song2019}
Song M~Y, Yoon J~S, Cho H, Karwasz G~P, Kokoouline V, Nakamura Y and Tennyson J
  2019 {\em Journal of Physical and Chemical Reference Data\/} {\bf 48} 043104
  ISSN 0047-2689 \urlprefix\url{https://doi.org/10.1063/1.5114722}

\bibitem{Brunger2017}
Brunger M~J 2017 {\em International Reviews in Physical Chemistry\/} {\bf 36}
  333--376 ISSN 1366591X
  \urlprefix\url{https://doi.org/10.1080/0144235X.2017.1301030}

\bibitem{Sanz2012}
Sanz A~G, Fuss M~C, Mu{\~{n}}oz A, Blanco F, Lim{\~{a}}o-Vieira P, Brunger M~J,
  Buckman S~J and Garc{\'{i}}a G 2012 {\em International Journal of Radiation
  Biology\/} {\bf 88} 71--76 ISSN 0955-3002
  \urlprefix\url{https://doi.org/10.3109/09553002.2011.624151}

\bibitem{Brunger2016}
Brunger M~J, Ratnavelu K, Buckman S~J, Jones D~B, Mu{\~{n}}oz A, Blanco F and
  Garc{\'{i}}a G 2016 {\em The European Physical Journal D\/} {\bf 70} 46 ISSN
  1434-6060 \urlprefix\url{https://doi.org/10.1140/epjd/e2016-60641-8}

\bibitem{White2018}
White R~D, Cocks D, Boyle G, Casey M, Garland N, Konovalov D, Philippa B,
  Stokes P, {De Urquijo} J, Gonz{\'{a}}lez-Magaa O, McEachran R~P, Buckman S~J,
  Brunger M~J, Garcia G, Dujko S and Petrovic Z~L 2018 {\em Plasma Sources
  Science and Technology\/} {\bf 27} 053001 ISSN 13616595
  \urlprefix\url{https://doi.org/10.1088/1361-6595/aabdd7}

\bibitem{Takeuchi2001}
Takeuchi T and Nakamura Y 2001 {\em IEEJ Transactions on Fundamentals and
  Materials\/} {\bf 121} 481--486 ISSN 0385-4205
  \urlprefix\url{https://doi.org/10.1541/ieejfms1990.121.5%5F481}

\bibitem{Robson2011}
Robson R~E, White R~D and Ness K~F 2011 {\em The Journal of Chemical Physics\/}
  {\bf 134} 064319 ISSN 0021-9606
  \urlprefix\url{https://doi.org/10.1063/1.3544210}

\bibitem{Kondo1990}
Kondo K and Tagashira H 1990 {\em Journal of Physics D: Applied Physics\/} {\bf
  23} 1175--1183 ISSN 0022-3727
  \urlprefix\url{https://doi.org/10.1088/0022-3727/23/9/007}

\bibitem{Brunger2000}
Brunger M~J, Campbell L, Cartwright D~C, Middleton A~G, Mojarrabi B and Teubner
  P~J~O 2000 {\em Journal of Physics B: Atomic, Molecular and Optical
  Physics\/} {\bf 33} 809--819 ISSN 0953-4075
  \urlprefix\url{https://doi.org/10.1088/0953-4075/33/4/315}

\bibitem{Cartwright2000}
Cartwright D~C, Brunger M~J, Campbell L, Mojarrabi B and Teubner P~J~O 2000
  {\em Journal of Geophysical Research: Space Physics\/} {\bf 105} 20857--20867
  ISSN 01480227 \urlprefix\url{https://doi.org/10.1029/1999JA000333}

\bibitem{Campbell1997}
Campbell L, Teubner P~J~O, Brunger M~J, Mojarrabi B and Cartwright D~C 1997
  {\em Australian Journal of Physics\/} {\bf 50} 525 ISSN 0004-9506
  \urlprefix\url{https://doi.org/10.1071/P96060}

\bibitem{Xu2018}
Xu X, Xu L~Q, Xiong T, Chen T, Liu Y~W and Zhu L~F 2018 {\em The Journal of
  Chemical Physics\/} {\bf 148} 044311 ISSN 0021-9606
  \urlprefix\url{https://doi.org/10.1063/1.5019284}

\bibitem{Kato2007}
Kato H, Kawahara H, Hoshino M, Tanaka H and Brunger M 2007 {\em Chemical
  Physics Letters\/} {\bf 444} 34--38 ISSN 00092614
  \urlprefix\url{https://doi.org/10.1016/j.cplett.2007.06.134}

\bibitem{Campbell2004}
Campbell L, Brunger M~J, Petrovic Z~L, Jelisavcic M, Panajotovic R and Buckman
  S~J 2004 {\em Geophysical Research Letters\/} {\bf 31} n/a--n/a ISSN 00948276
  \urlprefix\url{https://doi.org/10.1029/2003GL019151}

\bibitem{Josic2001}
Josi{\'{c}} L, Wr{\'{o}}blewski T, Petrovi{\'{c}} Z, Mechli{\'{n}}ska-Drewko J
  and Karwasz G 2001 {\em Chemical Physics Letters\/} {\bf 350} 318--324 ISSN
  00092614 \urlprefix\url{https://doi.org/10.1016/S0009-2614(01)01310-0}

\bibitem{Jelisavcic2003}
Jelisavcic M, Panajotovic R and Buckman S~J 2003 {\em Physical Review
  Letters\/} {\bf 90} 203201 ISSN 0031-9007
  \urlprefix\url{https://doi.org/10.1103/PhysRevLett.90.203201}

\bibitem{Mojarrabi1995}
Mojarrabi B, Gulley R~J, Middleton A~G, Cartwright D~C, Teubner P~J~O, Buckman
  S~J and Brunger M~J 1995 {\em Journal of Physics B: Atomic, Molecular and
  Optical Physics\/} {\bf 28} 487--504 ISSN 0953-4075
  \urlprefix\url{https://doi.org/10.1088/0953-4075/28/3/019}

\bibitem{Rapp1965}
Rapp D and Briglia D~D 1965 {\em The Journal of Chemical Physics\/} {\bf 43}
  1480--1489 ISSN 0021-9606 \urlprefix\url{https://doi.org/10.1063/1.1696958}

\bibitem{Lindsay2000}
Lindsay B~G, Mangan M~A, Straub H~C and Stebbings R~F 2000 {\em The Journal of
  Chemical Physics\/} {\bf 112} 9404--9410 ISSN 0021-9606
  \urlprefix\url{https://doi.org/10.1063/1.481559}

\bibitem{Alle1996}
Alle D~T, Brennan M~J and Buckman S~J 1996 {\em Journal of Physics B: Atomic,
  Molecular and Optical Physics\/} {\bf 29} L277--L282 ISSN 0953-4075
  \urlprefix\url{https://doi.org/10.1088/0953-4075/29/7/006}

\bibitem{Brunger2017a}
Brunger M~J, Buckman S~J and Ratnavelu K 2017 {\em Journal of Physical and
  Chemical Reference Data\/} {\bf 46} 023102 ISSN 0047-2689
  \urlprefix\url{https://doi.org/10.1063/1.4982827}

\bibitem{Boyle2017}
Boyle G~J, Tattersall W~J, Cocks D~G, McEachran R~P and White R~D 2017 {\em
  Plasma Sources Science and Technology\/} {\bf 26} 024007 ISSN 13616595
  (\textit{Preprint} \eprint{1509.00867})
  \urlprefix\url{https://doi.org/10.1088/1361-6595/aa51ef}

\bibitem{White2003}
White R~D, Robson R~E, Schmidt B and Morrison M~A 2003 {\em Journal of Physics
  D: Applied Physics\/} {\bf 36} 3125--3131 ISSN 00223727
  \urlprefix\url{https://doi.org/10.1088/0022-3727/36/24/006}

\bibitem{Skinker1923}
Skinker M and White J 1923 {\em The London, Edinburgh, and Dublin Philosophical
  Magazine and Journal of Science\/} {\bf 46} 630--637 ISSN 1941-5982
  \urlprefix\url{https://doi.org/10.1080/14786442308634289}

\bibitem{Bailey1934}
Bailey V and Somerville J 1934 {\em The London, Edinburgh, and Dublin
  Philosophical Magazine and Journal of Science\/} {\bf 17} 1169--1176 ISSN
  1941-5982 \urlprefix\url{https://doi.org/10.1080/14786443409462470}

\bibitem{Townsend1948}
Townsend J~S 1948 {\em {Motion of Slow Electrons in Gases}\/} (London:
  Hutchinson)

\bibitem{Parkes1972}
Parkes D~A and Sugden T~M 1972 {\em Journal of the Chemical Society, Faraday
  Transactions 2\/} {\bf 68} 600 ISSN 0300-9238
  \urlprefix\url{https://doi.org/10.1039/F29726800600}

\bibitem{Lakshminarasimha1977}
Lakshminarasimha C~S and Lucas J 1977 {\em Journal of Physics D: Applied
  Physics\/} {\bf 10} 313--321 ISSN 0022-3727
  \urlprefix\url{https://doi.org/10.1088/0022-3727/10/3/011}

\bibitem{Mechlinska-Drewko1999}
Mechlinska-Drewko J, Roznerski W, Petrovic Z~L and Karwasz G~P 1999 {\em
  Journal of Physics D: Applied Physics\/} {\bf 32} 2746--2749 ISSN 0022-3727
  \urlprefix\url{https://doi.org/10.1088/0022-3727/32/21/306}

\bibitem{Casey2020}
Casey M~J, Stokes P~W, Cocks D~G, Bo{\v{s}}njakovi{\'{c}} D, Simonovi{\'{c}} I,
  Brunger M~J, Dujko S, Petrovi{\'{c}} Z~L, Robson R~E and White R~D 2021 {\em
  Plasma Sources Science and Technology\/} {\bf 30} ISSN 13616595

\bibitem{Townsend1913}
Townsend J~S~E and Tizard H~T 1913 {\em Proceedings of the Royal Society of
  London. Series A, Containing Papers of a Mathematical and Physical
  Character\/} {\bf 88} 336--347 ISSN 0950-1207
  \urlprefix\url{https://royalsocietypublishing.org/doi/10.1098/rspa.1913.0034}

\bibitem{Bradbury1934}
Bradbury N~E 1934 {\em The Journal of Chemical Physics\/} {\bf 2} 827--834 ISSN
  0021-9606 \urlprefix\url{https://doi.org/10.1063/1.1749403}

\bibitem{Biagi}
{Biagi database} \urlprefix\url{www.lxcat.net/Biagi}

\bibitem{Biagib}
Biagi S {MAGBOLTZ}

\bibitem{LXCat}
{LXCat} \urlprefix\url{www.lxcat.net}

\bibitem{Misra2019}
Misra D 2019  (\textit{Preprint} \eprint{1908.08681})
  \urlprefix\url{http://arxiv.org/abs/1908.08681}

\bibitem{White2009}
White R~D, Robson R~E, Dujko S, Nicoletopoulos P and Li B 2009 {\em Journal of
  Physics D: Applied Physics\/} {\bf 42} 194001 ISSN 00223727
  \urlprefix\url{https://doi.org/10.1088/0022-3727/42/19/194001}

\bibitem{Innes2018}
Innes M 2018 {\em Journal of Open Source Software\/} {\bf 3} 602 ISSN 2475-9066
  \urlprefix\url{https://doi.org/10.21105/joss.00602}

\bibitem{Glorot2010}
Glorot X and Bengio Y 2010 {Understanding the difficulty of training deep
  feedforward neural networks} {\em Journal of Machine Learning Research\/}
  ({\em Proceedings of Machine Learning Research\/} vol~9) ed Teh Y~W and
  Titterington M (Chia Laguna Resort, Sardinia, Italy: PMLR) pp 249--256 ISSN
  15324435 \urlprefix\url{http://proceedings.mlr.press/v9/glorot10a.html}

\bibitem{zhuang2020adabelief}
Zhuang J, Tang T, Ding Y, Tatikonda S~C, Dvornek N, Papademetris X and Duncan J
  2020 {\em Advances in Neural Information Processing Systems\/} {\bf 33}

\bibitem{Nesterov1983}
Nesterov Y 1983 {\em Soviet Mathematics Doklady\/} {\bf 27} 372--376

\bibitem{Dozat2016}
Dozat T 2016 {Incorporating Nesterov Momentum into Adam} {\em Proceedings of
  4th International Conference on Learning Representations, Workshop Track\/}

\bibitem{Bishop1994}
Bishop C~M 1994 {Mixture density networks}
  \urlprefix\url{http://publications.aston.ac.uk/id/eprint/373/}

\bibitem{Sohn2015}
Sohn K, Lee H and Yan X 2015 {Learning Structured Output Representation using
  Deep Conditional Generative Models} {\em Advances in Neural Information
  Processing Systems 28\/} ed Cortes C, Lawrence N~D, Lee D~D, Sugiyama M and
  Garnett R (Curran Associates, Inc.) pp 3483--3491
  \urlprefix\url{http://papers.nips.cc/paper/5775-learning-structured-output-representation-using-deep-conditional-generative-models.pdf}

\bibitem{Mirza2014}
Mirza M and Osindero S 2014  (\textit{Preprint} \eprint{1411.1784})
  \urlprefix\url{http://arxiv.org/abs/1411.1784}

\bibitem{Dinh2014}
Dinh L, Krueger D and Bengio Y 2015 {\em 3rd International Conference on
  Learning Representations, ICLR 2015 - Workshop Track Proceedings\/}
  (\textit{Preprint} \eprint{1410.8516})
  \urlprefix\url{https://arxiv.org/abs/1410.8516}

\bibitem{Dinh2016}
Dinh L, Sohl-Dickstein J and Bengio S 2017 {\em 5th International Conference on
  Learning Representations, ICLR 2017 - Conference Track Proceedings\/}
  (\textit{Preprint} \eprint{1605.08803})
  \urlprefix\url{https://arxiv.org/abs/1605.08803}

\bibitem{Kingma2018}
Kingma D~P and Dhariwal P 2018 {Glow: Generative flow with invertible 1×1
  convolutions} {\em Advances in Neural Information Processing Systems\/} vol
  2018-Decem ed Bengio S, Wallach H, Larochelle H, Grauman K, Cesa-Bianchi N
  and Garnett R (Curran Associates, Inc.) pp 10215--10224 (\textit{Preprint}
  \eprint{1807.03039})
  \urlprefix\url{http://papers.nips.cc/paper/8224-glow-generative-flow-with-invertible-1x1-convolutions.pdf}

\bibitem{White2014}
White R~D, Brunger M~J, Garland N~A, Robson R~E, Ness K~F, Garcia G, {De
  Urquijo} J, Dujko S and Petrovi{\'{c}} Z~L 2014 {\em European Physical
  Journal D\/} {\bf 68} 125 ISSN 14346079
  \urlprefix\url{https://doi.org/10.1140/epjd/e2014-50085-7}

\end{thebibliography}

\end{document}